\documentclass[twocolumn,traditabstract]{aa}
\usepackage{txfonts}
\usepackage{color}
\usepackage{graphicx}
\usepackage{natbib}
\usepackage{multirow}
\begin{document}

  \title{Herschel/PACS view of disks around low-mass stars and brown dwarfs in the TW Hya association 
  \thanks{{\it Herschel} is an ESA space observatory with science instruments provided by European-led Principal Investigator 
   consortia and with important participation from NASA.}}  
   \author{Yao Liu$^{1,2}$
         \and
         Gregory J. Herczeg$^{3}$
         \and
         Munan Gong$^{4,5}$
         \and
         Katelyn N. Allers$^{6}$
         \and
         Joanna M. Brown$^{7}$
         \and
         Adam L. Kraus$^{8}$ 
         \and
         Michael C. Liu$^{9}$
         \and
         Evgenya L. Shkolnik$^{10}$
         \and
         Ewine F. van Dishoeck$^{11, 12}$}
  \institute{Purple Mountain Observatory, Chinese Academy of Sciences, 2 West Beijing Road, Nanjing 210008, China
            \and
            Key Laboratory for Radio Astronomy, Chinese Academy of Sciences, 2 West Beijing Road, Nanjing 210008, China
            \and
            Kavli Institute for Astronomy and Astrophysics, Peking University, Yi He Yuan Lu 5, Haidian Qu, 
            Beijing 100871, China
            \and
            Tsinghua University, Shuang Qing Lu 30, Haidian Qu, Beijing 100084, China
            \and
            Department of Astrophysical Sciences, Princeton University, 4 Ivy Lane, Peyton Hall, Princeton, NJ 08544, USA
            \and
            Department of Physics and Astronomy, Bucknell University, Lewisburg, PA 17837, USA            
            \and
            Harvard-Smithsonian Center for Astrophysics, 60 Garden St., MS 78, Cambridge, MA 02138, USA            
            \and
            Department of Astronomy, The University of Texas at Austin, Austin, TX 78712, USA           
            \and
            Institute for Astronomy, University of Hawaii at Manoa, 2680 Woodlawn Dr., Honolulu, HI 96822, USA           
            \and
           Lowell Observatory, 1400 West Mars Hill Road, Flagstaff, AZ, 86001, USA            
            \and
            Leiden Observatory, Leiden University, PO Box 9513, 2300 RA Leiden, The Netherlands 
            \and
           Max-Planck-Institut f$\ddot{\rm{u}}$r Extraterrestrische Physik, Giessenbachstrasse 1, 85748 Garching, Germany}
  \authorrunning{Liu et al.}
  \titlerunning{PACS photometry of very low mass stars and brown dwarfs in the TWA}

\abstract{We conducted {\it Herschel}/PACS observations of five very low-mass stars or brown dwarfs located in the
TW Hya association with the goal of characterizing the properties of disks in the low stellar mass regime.
We detected all five targets at $70\,\mu{\rm{m}}$ and $100\,\mu{\rm{m}}$ and three targets at $160\,\mu{\rm{m}}$.
Our observations, combined with previous photometry from 2MASS, WISE, and SCUBA-2, enabled us to construct 
SEDs with extended wavelength coverage. Using sophisticated radiative transfer models, we analyzed the observed SEDs of the 
five detected objects with a hybrid fitting strategy that combines the model grids and the simulated annealing algorithm and 
evaluated the constraints on the disk properties via the Bayesian inference method. The modeling suggests that disks 
around low-mass stars and brown dwarfs are generally flatter than their higher mass counterparts, but the range of disk mass 
extends to well below the value found in T Tauri stars, and the disk scale heights are comparable in both groups.  
The inferred disk properties (i.e., disk mass, flaring, and scale height) in the low stellar mass regime are consistent 
with previous findings from large samples of brown dwarfs and very low-mass stars. We discuss the dependence of disk 
properties on their host stellar parameters and find a significant correlation between the {\it Herschel} far-IR fluxes 
and the stellar effective temperatures, probably indicating that the scaling between the stellar and disk 
masses (i.e., $M_{\rm{disk}} \propto M_{\star}$) observed mainly in low-mass stars may extend down to the brown dwarf regime.}


\maketitle

\section{Introduction}
\label{sec:intro}
Circumstellar disks form as a natural result of angular momentum conservation during the early phase 
of star formation, i.e., collapse of rotating protostellar cores \citep{shu1987}. They gradually lose 
mass and eventually dissipate through complex mechanisms such as accretion, photo-evaporation, and planet 
formation \citep[e.g.,][]{alexander2006, balog2008, williams2011}. How brown dwarfs and very low-mass 
stars form is an interesting question in the field of star formation and still remains a subject of 
debate although several theories have been proposed, for instance, a scaled down version of star formation 
processes, gravitational instabilities in disks and ejection of the stellar 
embryo \citep[e.g.,][]{luhman2012, chabrier2014}. The disk properties of the (sub)stellar objects play a 
crucial role in understanding their formation. Numerous observations of brown dwarfs in recent years have 
detected disks and envelopes with properties similar to those found in T Tauri stars, suggesting that brown 
dwarfs may form in a similar way to hydrogen-burning 
stars \citep[e.g.,][]{liun2003, white2003, apai2005, scholz2006, andre2012, harvey2012a, harvey2012b, joergens2013}. 
Despite this progress, more thorough disk comparisons between the properties 
of disks in different stellar mass regimes are required to obtain a clear view of the formation mechanism 
of brown dwarfs and very low-mass stars.
 
For comparative studies between clusters, one of the foundational regions is the nearby TW Hya
Association \citep[TWA,][]{webbz1999}. Comparing the stellar temperatures and luminosities of TWA 
members that have been identified to pre-main sequence tracks yields ages of  ${\sim} 7-10$ Myr 
\citep{weinberger2013, ducourantt2014}, young enough that ${\sim} 20\%$ of the primordial disks still exist but old enough 
that the surviving disks have likely undergone significant evolution. The membership of TWA still remains incomplete. 
Given its relatively old age and proximity to the earth, the TWA is an ideal place for studying the 
properties of disks and disk evolution in the low stellar mass regime \citep{looperb2010, schneiders2012}.

\begin{table}[!t]
\caption{\bf Observation log.}
\centering
\begin{tabular}{lcccccc}
\hline
Target       &  Date & AOR    &   $\lambda$ ($\mu$m) &  $t_{{\rm exp}}^a$  (s) \\
\hline
TWA 30A  & 2011-11-28  & 1342233252-53 & 70,160 & 1340\\
TWA 30A  & 2011-11-28  & 1342233254-55 & 100,160 & 1114\\
TWA 30B  & \multicolumn{3}{c}{Same images as TWA 30A}  \\
TWA 31    & 2011-07-14 & 1342224190-91 & 70,160 & 890\\
TWA 31    & 2011-07-14  & 1342224192-93 & 100,160 & 440\\
TWA 31    & 2011-12-16 & 1342234393, 96 & 70,160 & 1292\\
TWA 31    & 2011-12-16  & 1342234394-95 & 100,160 & 858\\
TWA 32   &  2011-07-28 & 1342224916-17 & 70,160 & 890\\
TWA 32   &  2011-07-28 & 1342224918-19 & 70,160 & 440\\
TWA 34  & 2011-11-27 & 1342233099-100 & 70,160 & 890\\
TWA 34  & 2011-11-27 & 1342233100-101 & 100,160 & 440\\
\hline
\multicolumn{5}{l}{$^a$total exposure time for both orientations}\\
\hline
\end{tabular}
\label{tab:obslog}
\end{table}

%\begin{table*}[!t]
%\caption{Target properties.}
%\centering
%\begin{tabular}{lccccc}
%\hline
% Target & $d$  & SpT & $T_{\rm{eff}}$ & $\log L$ & H$\alpha$ EW    \\
%        & [pc] &   & [K] & [$L/L_\odot$] & [$\AA$]     \\
%  \hline
%TWA 30A & $56^{+7}_{-7}$    & M5   & 2980 & -1.29 & -6.8    \\
%TWA 30B & $56^{+21}_{-12}$  & M4   & 3190 & -1.2  & -7.4     \\
%TWA 31  & 110               & M4.2 & 3150 & -2.0  & -115     \\
%TWA 32  & $77^{+4}_{-4}$    & M6.3 & 2830 & -1.70 & -12.6    \\
%TWA 34  & $50^{+4}_{-4}$    & M4.9 & 3000 & -1.83 & -9.6     \\
%\hline
%\end{tabular}
%\tablefoot{The distance to each target, spectral type (SpT), effective temperature, luminosity and the accretion properties are described 
%           in Section 2.1.}
%\label{tab:properties}
%\end{table*}

\begin{table*}[!t]
\caption{Target properties and IR photometry of detected TWA disks.}
\centering
\begin{tabular}{lcccccccccccc}
\hline
 Target& $d$  & SpT & $T_{\rm{eff}}$ & $\log L$ & H$\alpha$ EW  & W1 & W2 & W3 & W4 & PACS 70 & PACS 100 & PACS 160 \\
       & (pc) &   & (K) & ($L/L_\odot$) & ($\AA$)  & (mJy) & (mJy) & (mJy) & (mJy) & (mJy) & (mJy) & (mJy)  \\
  \hline
TWA 30A & 56 $\pm$ 7        & M5   & 2980 & $-1.29\pm0.12$ & $-$6.8  & 93.0 & 72.1 & 43.1  & 66.0  & $15\pm1.2$   & $10.5\pm1.3$ & $1.4\pm1.8$ \\
TWA 30B & $56^{+21}_{-12}$  & M4   & 3190 & $-1.2\pm0.2$   & $-$7.4  & 3.68 & 9.63 & 33.50 & 71.45 & $65.7\pm1.8$ & $55.6\pm2.1$ & $48.4\pm3.0$\\
TWA 31  & 110               & M4.2 & 3150 & $-2.0$         & $-$115  & 6.11 & 5.29 & 3.05  & 5.59  & $9.1\pm0.9$  & $7.1\pm0.9$  & $3.4\pm2.1$ \\
TWA 32  & 77 $\pm$ 4        & M6.3 & 2830 & $-1.70\pm0.12$ & $-$12.6 & 46.2 & 35.4 & 22.3  & 36.9  & $46.9\pm1.3$ & $51.4\pm2.3$ & $46.9\pm2.0$ \\
TWA 34  & 50 $\pm$ 4        & M4.9 & 3000 & $-1.83\pm0.14$ & $-$9.6  & 44.7 & 38.5 & 14.7  & 13.9  & $24.5\pm1.2$ & $18.7\pm1.8$ & $17.2\pm2.5$ \\
\hline
\end{tabular}
\tablefoot{(1) The distance to each target, spectral type (SpT), effective temperature, luminosity, and the accretion properties are described 
            in Sect. 2.1. (2) W1, W2, W3, and W4 refer to the photometry taken with WISE at 3.4, 4.6, 12, and $22\,\mu{\rm{m,}}$ respectively \citep{cutri2012}.}
\label{tab:photometry}
\end{table*}

The mass is one of the most important parameters of a disk because it sets the critical condition
of whether giant planets can form within the disk.
Most measurements of disk masses have been obtained from the (sub-)millimeter (mm) range
\citep[e.g.,][]{andrewsw2005, andrewsr2013, mohantyg2013}. Since
massive disks are almost optically thin at (sub)-mm wavelengths, the measured continuum flux can
be directly converted into a disk mass, with a relative accuracy
that is dependent on a similarity in grain growth, composition, and disk structure, as well as on the gas-to-dust mass ratio.
For very low-mass stars and brown dwarfs, there are only a few measurements of 
disk masses because of the sensitivity limits of ground-based sub-mm continuum observations.
However, far-infrared (far-IR) wavelengths accessible by {\it Herschel}/PACS have been successfully used to 
estimate disk masses of very low-mass stars and brown dwarfs because most parts of the disk in this case 
become optically thin in the far-IR and are much brighter in the far-IR than in the sub-mm \citep{harvey2012a, harvey2012b}. 
Moreover, far-IR measurements supplemented with near- and mid-IR observations 
provide additional constraints on the structure of circumstellar disks.

In this paper, we focus on the PACS detections of five disk-bearing low-mass stars and 
brown dwarfs in the TWA. Our goal is to characterize the disk properties around 
these low stellar mass objects and try to compare the structure of disks in different 
stellar mass regimes. 

%The paper is organized as follows. We present the PACS observations and data 
%processing in the following section. A detailed description of the SED modeling process 
%is described in Section 3. We discuss our results in Section 4, followed by a
%brief summary in Section 5.

\section{Sample and observations}

\subsection{Sample properties}
\label{sec:sample}

Our sample consists of five low-mass members of the TWA that do not have existing 
far-IR photometry, i.e., TWA 30A \citep{looperm2010}, TWA 30B \citep{looperb2010}, TWA 31, TWA 32 \citep{shkolnikl2011}, and 
TWA 34 \citep{schneiders2012}. Object properties are described in Table~\ref{tab:photometry}. The spectral types of our sample 
lie in the range of [M4, $\sim$M6.5], corresponding to very low-mass stars or brown dwarfs according to theoretical 
evolutionary models \citep{chabrier2000}. We derive the (sub)stellar temperatures using the spectral type to 
temperature conversion presented in \citet{herczeg2014}.  For TWA 31, we adopted a distance of 110 pc estimated 
by \citet{shkolnikl2011} under an assumption of its membership in the TWA.
Distances for the four other objects are obtained from dynamical modeling of the TWA by \citet{ducourantt2014}. 
We fit the photospheric emission of each target using the grid of BT-Settl models \citep{allardh2012} with 
corresponding temperature and distance, thereby determining the (sub)stellar luminosity from the best-fit model 
deduced with a $\chi^2$ minimization. In this procedure, we assume solar metallicity, a gravity of ${\rm log}\,g=3.5,$ 
and no interstellar extinction for all the targets. The fluxes at the 2MASS wavelengths are assumed to follow the 
phototsphere. In some cases such as TWA 30A, we also presume no excess in the WISE $3.4\,\mu{\rm{m}}$ band. 
The uncertainty in luminosity is dominated by errors in the dynamical distance calculations, which are 
typically quoted as $\sim 10\%$, see Table \ref{tab:photometry}.

TWA 32 is a $0\farcs6$, near-equal mass binary \citep{shkolnikl2011}. In our analysis, we assume that only one of 
the two stars in the system retains a disk, similar to the Hen 3-600 and HD 98800 multiple star systems in the TWA \citep{andrews2010}.
The optical/near-IR photometry used for SED modeling (see Sect. \ref{sec:modeling}) therefore should be adjusted to 
properly eliminate the contribution from the binary companion. We first fit the photospheric emission of the whole 
system, then altered the photometry by subtracting half of the best-fit photospheric flux at each wavelength. 

Rich emission line spectra from TWA 30A and TWA 30B and strong H$\alpha$ emission from TWA 31 indicate ongoing 
accretion \citep{looperb2010,looperm2010,shkolnikl2011}. The H$\alpha$ line equivalent widths for TWA 32 and 
TWA 34 \citep{shkolnikl2011,schneiders2012} are consistent with expectations for the chromospheric activity of 
young mid-M dwarfs \citep{white2003}, although in some cases accretion may not produce bright H$\alpha$ emission. 
\citet{looperb2010} found that the near-IR excess of TWA 30B is highly variable over timescales of a day. 
They suggested a highly inclined disk around this interesting object that causes changes in extinction induced 
by spatial variations in the disk structure rotating into our line of sight and out of it. The stellar luminosity 
cannot be empirically calculated from optical or near-IR photometry in this case because of obscuration by the 
nearly edge-on disk. Therefore, we infer a luminosity of $\log L/L_\odot=-1.2\pm0.2$ for TWA 30B based on 
the luminosities of other stars in the TWA with a similar spectral type, as calculated in this paper and by \citet{herczeg2014}.

\subsection{Observational setup}

We used {\it Herschel}/PACS \citep{pilbrattr2010, poglitschw2010} to obtain far-IR photometry of five
low-mass members of the TWA. The light in PACS is split into blue and red channels by a dichroic filter at
$\sim$120 $\mu$m. The integration time in the blue channel was split evenly between imaging with the 70 
and 100 $\mu$m (see $t_{\rm{exp}}$ in Table \ref{tab:obslog}). The red channel images photons with a 
central wavelength of $160\,\mu{\rm{m}}$.

The PACS images were obtained with eight scan legs of $3^\prime$ each with
steps of $4^{\prime\prime}$ at orientation angles of 70 and 110$^\circ$.  The
spatial resolution was about $5^{\prime\prime}$ at $70\,\mu{\rm{m}}$
and $10^{\prime\prime}$ at $160\,\mu{\rm{m}}$. The data were processed in the
Herschel Interactive Processing Environment (HIPE, version 6.0) with 
standard reduction routines.

\subsection{Source flux determination}

Fluxes are extracted from apertures of two pixels in radius, and subsequently corrected for the 
encircled energy fraction (0.55, 0.47, and 0.44 at 70, 100, and 160 $\mu$m) within the extraction
region. The background was averaged in an annulus between
60--72$^{\prime\prime}$ from the target. The locations of the
detected emission are consistent with the object location to within
the ${\sim} 1-2^{\prime\prime}$ pointing accuracy of the observations.  

The photometric flux densities and their 1$\sigma$ errors of the five detected
targets are summarized in Table \ref{tab:photometry}. The absolute flux
calibration is accurate to ${\sim} 5$\%. The typical offset from the nominal
position, ${\sim} 2''$, corresponding to ${\sim} 1$ pixel in the image, is
essentially identical in all wavelength channels and is well within
the typical {\it Herschel} pointing uncertainty.

\begin{figure*}[!t]
  \centering
   \includegraphics[width=\textwidth]{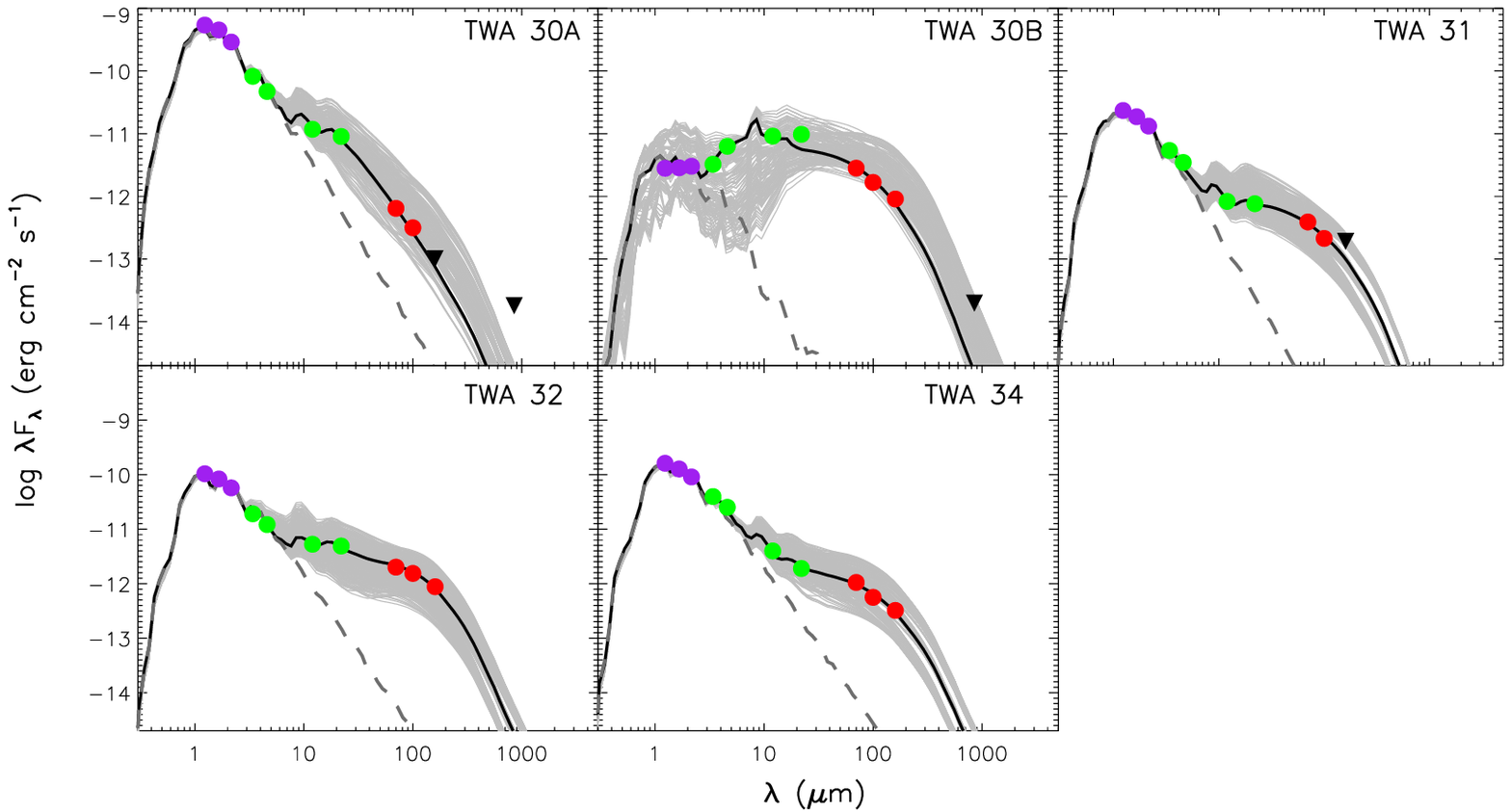}   
   \caption{Spectral energy distributions of the brown dwarf disks. The dots depict the photometry
            taken with 2MASS (purple), WISE (green), and {\it Herschel}/PACS (red). The upside down triangles 
            show the $3\,\sigma$ upper limits measured with PACS and SCUBA-2. The best-fit models are indicated 
            as black solid lines, whereas the dashed lines represent the photospheric emission levels. 
            The gray lines denote all the models that are within the uncertainties given in Table \ref{tab:bestfit},
            showing the constraints of the best fit models for each object. To deduce the uncertainties of model parameters 
            with Bayesian analysis, we only take the models calculated in the grid into account, see Sect. \ref{sec:modeling}. 
            Notes: (1) for the binary TWA 32, the near-IR photometry shown in this figure are $F_{\rm{2MASS\,J}}=42.6\,\rm{mJy}$, 
            $F_{\rm{2MASS\,H}}=46.1\,\rm{mJy}$ and $F_{\rm{2MASS\,K}}=41.2\,\rm{mJy}$. These values are different from 
            the observed ones of the whole binary system, see Sect. \ref{sec:sample}. (2) For TWA 30B, all the models indicated 
            as gray lines feature $i=75^{\circ}$. Models with other inclinations are not shown because the derivation of uncertainties 
            in $R_{\rm{in}}$, $M_{\rm{disk}}$, $\beta$ and $h_{100}$ is based on models with $i=75^{\circ}$ in the grid; 
            see Sect. \ref{sec:modeling} for an explanation.}
\label{fig:bestfit}
\end{figure*}

\begin{table*}[t]
\caption{Validity ranges from the Bayesian analysis for the disk parameters.}
\centering
\begin{tabular}{lcccccccc}
\hline
 Object      & $R_{\rm{in}}$ & $R_{\rm{out}}$ & ${\rm log}_{10}\,M_{\rm{disk}}$      & $\beta$ & $h_{100}$ & Inclination \\
             &  (AU)         &    (AU)        & ($M_{\odot}$) &         &    (AU)     & ($^{\circ}$) \\
  \hline
 TWA 30A    & 0.15 (0.023, 0.218)  &100  &$-$6.0 ($-$6, $-$5)       &1.01 (1.0, 1.1)   &2 (2, 5)      &45 (15, 75)    \\ 
 TWA 30B    &0.079 (0.018, 0.254)  &100  &$-$5.25 ($-$5.5, $-$4)    &1.02 (1.0, 1.1)   &12.6 (8, 17)  &83 (60, 90)    \\
 TWA 31     &0.008 (0.007, 0.144)  &100  &$-$5.5 ($-$6, $-$4.75)    &1.20 (1.1, 1.25)  &20 (11, 20)   &30 (15, 67.5)  \\
 TWA 32     &0.133 (0.029, 0.603)  &30   &$-$5.0 ($-$5.75, $-$4.5)  &1.15 (1.1, 1.25)  &9 (5, 14)     &15 (15, 75)    \\
 TWA 34     &0.009 (0.009, 0.039)  &100  &$-$5.5 ($-$6, $-$5)       &1.15 (1.1, 1.25)  &11 (5, 14)    &60 (15, 67.5)  \\
 \hline
\end{tabular}
\tablefoot{The first number of each parameter corresponds to the best-fit value. The values quoted in brackets give the 
ranges of validity for each parameter deduced from Bayesian analysis. The total disk mass $M_{\rm{disk}}$ is calculated from the
dust mass assuming a gas-to-dust mass ratio of 100.}
\label{tab:bestfit}
\end{table*}

\section{SED Modeling}
\label{sec:modeling}
All five objects have photometry from the Two Micron All Sky Survey \citep[2MASS,][]{skrutskie2006} and 
the Wide-Field Infrared Explorer \citep[WISE,][]{wright2010} surveys. For TWA 30A and TWA 30B, \citet{mohantyg2013} 
also derived upper limits of the flux in the sub-mm from SCUBA-2. We modeled the broadband SEDs of our targets 
using the radiative transfer code \texttt{MC3D} developed by \citet{wolf2003a} in order to characterize the structure 
of their surrounding disks. 
~~~~~\\ ~~~~~\\
\noindent{\it Dust distribution in the disk:}\hspace*{2mm}
We introduce a parametrized flared disk in which dust and gas are well mixed and homogeneous throughout the system.
This model has been successfully used to explain the observed SEDs of a large sample of young stellar objects and 
brown dwarfs \citep[e.g.,][]{wolfp2003, sauter2009, harvey2012a}.
For the dust in the disk, we assume a density structure with a Gaussian vertical profile
\begin{equation}
\rho_{\rm{dust}}=\rho_{0}\left(\frac{R_{*}}{\varpi}\right)^{\alpha}\exp\left[-\frac{1}{2}\left(\frac{z}{h(\varpi)}\right)^2\right], \\
\label{dust_density}
\end{equation}
and a power-law distribution for the surface density \\
\begin{equation}
\Sigma(\varpi)=\Sigma_{0}\left(\frac{R_{*}}{\varpi}\right)^p,
\end{equation}
where $\varpi$ is the radial distance from the central star measured in the disk midplane, and $h(\varpi)$ is the scale 
height of the disk. The disk extends from an inner radius $R_{\rm{in}}$ to an outer radius $R_{\rm{out}}$. 
The $R_{\rm{out}}$ is fixed in the modeling process because the choice of this parameter makes essentially no difference 
to the synthetic SEDs in the simulated wavelength ranges \citep{harvey2012a}. For the binary TWA 32, we take a value 
of $30\,\rm{AU}$ that is expected to be less than the separation within the system. We set $R_{\rm{out}}=100\,\rm{AU}$ 
for the rest models. To allow flaring, the scale height follows the power law
\begin{equation}
h(\varpi) = h_{100}\left(\frac{\varpi}{100\,\rm{AU}}\right)^\beta,\\
\end{equation}
with the exponent $\beta$ describing the extent of flaring and the scale height $h_{100}$ at a distance of 
$100\,\rm{AU}$ from the central star. The indices $\alpha$, $p$, and $\beta$ are codependent, i.e., $p=\alpha-\beta$. 
To reduce the dimensionality of the parameter space, we fix $p=1$ in the simulation since pure SED modeling cannot place tight  
constraints on this parameter.
~~~~~\\ ~~~~~\\
\noindent{\it Dust properties:}\hspace*{2mm}
We consider the dust grains to be a homogeneous mixture of amorphous silicate and carbon with a mean 
density of $\rho_{\rm{grain}} = 2.5\,{\rm g\,cm^{-3}}$. The grain size distribution is given by the standard 
power law ${\rm d}n(a)\propto{a^{-3.5}} {\rm d}a$ with minimum and maximum grain 
sizes $a_{\rm{min}}=0.1\,\mu{\rm m}$ and $a_{\rm{max}}=100\,\mu{\rm m}$, respectively.
We set a low value for $a_{\rm{min}}$ to ensure that its exact value has a negligible impact on the 
synthetic SEDs. Since there is no information about the maximum grain size, such as the (sub-)mm spectral index,
in the five target disks, we increase $a_{\rm{max}}$ from the interstellar medium's value $0.25\,\mu{\rm m}$ \citep{mathis1977} 
to $100\,\mu{\rm m}$ to account for grain growth effects that are commonly observed in disks around T Tauri stars 
and brown dwarfs \citep[e.g.,][]{ricci2010, ricci2012, broekhoven2014}. With the Mie theory, we calculate the optical 
properties of each dust component using the complex refractive indices of 
amorphous silicate and carbon published by \citet{jager1994}, \citet{dorschner1995} and \citet{jager1998}.
Relative abundances of 75\% silicate and 25\% carbon are used to derive the weighted mean values of dust grain parameters,
for instance the absorption and scattering cross section \citep{wolf2003b}.
~~~~~\\ ~~~~~\\
\noindent{\it Heating sources:}\hspace*{2mm}
The disk is assumed to be passively heated by stellar irradiation \citep[e.g.,][]{chiangg1997}. 
We take the BT-Settl atmosphere models with ${\rm log}\,g=3.5$ as the incident substellar spectra \citep{allardh2012}. 
The radiative transfer problem is solved self-consistently considering 100 wavelengths, which are logarithmically distributed 
in the range of [$0.05\,\mu{\rm{m}}$, $2000\,\mu{\rm{m}}$].
~~~\\  ~~~\\
\noindent{\it Fitting method:}\hspace*{2mm} The task of SED fitting was performed with a hybrid strategy that combines 
the database method and the simulated annealing (SA) algorithm \citep{kirkpatrick1983}. A common method used to fit 
observational data is to precalculate a model database on a huge grid in parameter space. The fitting result can be 
identified very fast by evaluating the merit function (i.e., the $\chi^2$-distribution), once the database is built. 
The model grid can give us an overview of the fitting quality in different parameter domains through Bayesian analysis \citep{pinte2008}. 
However, because the number of grid points increases substantially with the dimensionality, the grid resolution always has 
to be coarse owing to finite computational resources and limited time. SA is a versatile optimization 
technique. Based on the Metropolis-Hastings algorithm, it creates a random walk, i.e. a Markov chain, through the parameter 
space, thereby gradually minimizing the discrepancy between observation and prediction by following the local 
topology of the merit function. This approach has specific advantages for high-dimensionality optimization because 
no gradients need to be calculated and local optimum can be overcome intrinsically regardless of the dimensionality. 
We first ran a large grid of disk models for each source considering a broad range of disk parameters. Then SA was used 
to improve the result by taking the best fit in the model grid as the starting point of the Markov chain. Moreover, the
initial step size of each dimension in the optimization process was set to be much smaller than the grid spacing. 
This kind of methodology makes use of the advantages of both database method and SA approach and has already been 
demonstrated to be successful for SED analysis \citep{lium2012, lium2013}. 

Table \ref{tab:sampling} summarizes the grid points of disk parameters calculated in the database. The grid adds up to a total number 
of 18~144 models for each target, whereas the length of a Markov chain is typically ${\sim}1000$ models when the SA algorithm
is aborted. The final fitting results are displayed in Fig. \ref{fig:bestfit}. The best-fit models are indicated as solid lines, whereas 
the dashed lines represent the photospheric emission level. The corresponding parameter sets are listed in Table \ref{tab:bestfit}. 
A comparison between the final best-fit parameter value and the grid points in the database can quantitatively evaluate
the improvements brought by SA to the modeling strategy. Taking TWA 32 as an example, the best-fit $h_{100,\rm{best-fit}}=9\,\rm{AU}$. 
The explored value of $h_{100}$ that is closest to $h_{100,\rm{best-fit}}$ in the database is $8\,\rm{AU}$ (see Table \ref{tab:sampling}),
therefore, the distance between the final fit and the best solution in the database is $\Delta\,h_{100}=1\,\rm{AU}$.
We notice that the adjustments introduced with SA to the best fits in the database for some targets (e.g., TWA 30A) are not too large,
indicating that databases can generally provide acceptable results. Cases like TWA 30B demonstrate that optimization with SA is 
indeed desired for increasing the fitting quality. We emphasize that the best fits presented in Fig. \ref{fig:bestfit} cannot be 
considered a unique solution since some of the model parameters are degenerate in the fitting process.
~~~\\  ~~~\\
\noindent{\it Bayesian analysis:}\hspace*{2mm} Despite the model degeneracy, previous studies have shown 
that modeling SEDs with broad wavelength coverage can constrain the mass and geometry of disks around brown 
dwarfs and low-mass stars \citep[e.g.,][]{harvey2012a, harvey2012b, olofsson2013, spezzi2013}. 
In particular, the Bayesian inference approach provides a statistical way to analyze the potential correlations 
and interplay between different parameters. The goodness of the fit is defined through a 
reduced $\chi_{\rm r}^2$. The relative probability of a given model is proportional to 
exp(-$\chi_{\rm r}^2/2$) in our case because we choose uniform a priori probabilities for each 
disk parameter. All probabilities are normalized at the end of the procedure so that the sum of 
the probabilities of all models is equal to 1. The probability distributions of each parameter in the 
range of values explored in the model grid are shown in several figures in the Appendix. 
With the results from Bayesian analysis, it is clear which parameters are well constrained, e.g., when the probability 
distribution is sharply concentrated within a relatively narrow range. In contrast, only loose constraints can be obtained 
if the probabilities of the parameter are flatly distributed. Table \ref{tab:bestfit} also gives 
the ranges (quoted in parenthesis) of validity for each parameter from its probability distribution, corresponding to 
regions where $P>0.5 \times P_{\rm Max}$. 

For TWA 30B, by using a luminosity of $\log L/L_\odot=-1.2\pm0.2$ (see Sect. \ref{sec:sample}) in the modeling, we 
implicitly introduce a prior probability distribution for the disk inclination (i.e., non-uniform), 
because all the models with $i<60^{\circ}$ will significantly overpredict the near- and mid-IR fluxes. 
Therefore, the Bayesian probability distributions for $R_{\rm{in}}$, $h_{100}$, $M_{\rm{disk}}$ and $\beta$ 
shown in Fig. \ref{fig:twa30b} are deduced from analyzing models with $i=75^{\circ}$ in the grid. The wide range of acceptable 
fits for TWA 30B shown in Fig. \ref{fig:bestfit} appear discrepant with the observed photometry. The optimal model found from 
the local fit with SA is far from any point in the initial model grid. However, even in the local fit, only 22 of 1000 sets of 
parameters yielded a $\chi^2$ below the confidence threshold, an unusually small number. The resulting optimal parameter space 
is very tight and likely not representative of the true uncertainties in the parameters, especially since the star is only detected 
in scattered light. The uncertainties obtained from the Bayesian analysis of the grid very likely yields more realistic error 
bars, despite some poor fits to the observed data within that range.

\section{Discussion}
\label{section:discussion}
We compiled observed SEDs of five low-mass stars or brown dwarfs in the TWA by combining our {\it Herschel}/PACS photometry
with previous data taken with 2MASS and WISE, and also for TWA 30A and TWA 30B with mm measurements. We exploited the broadband 
SEDs to characterize disk properties using sophisticated radiative transfer technique and a well-tested hybrid fitting 
approach. We also evaluated the constraints on different disk parameters through Bayesian analysis. 

\subsection{The disk properties of the targets}

The Bayesian probability distributions indicate poor constraints on the disk inclination for most objects.
Almost all the values sampled by our grid feature similar chances to reproduce the data
well. This is consistent with previous findings since the determination of this parameter
is based purely on the SED \citep[e.g.,][]{chiang1999, robitaille2007, harvey2012a}. Our results support 
the interpretation that the TWA 30B disk is viewed nearly edge-on. The quality of the SED fit drops 
significantly if the orientation is a small step outside the validity range presented here (i.e., $i<60^{\circ}$).

The disk inner radius is sensitive to the wavelength point at which the IR excess is dominated by 
the disk \citep{harvey2007}. Our model grid explores different values for this parameter logarithmically distributed in the 
range of [1, 200]$R_{\rm sub}$, where $R_{\rm sub}$ refers to the dust sublimation radius determined 
using the empirical relation $R_{\rm{sub}} = R_{\star}(T_{\rm{sub}}/T_{\rm{eff}})^{−2.085}$ from \citet{whitney2004}
for individual targets, and $T_{\rm{sub}}$ is the dust sublimation temperature and was set to $1600\,\rm{K}$.
Because the observed SEDs are well sampled in the near- and mid-IR domain, this parameter is relatively 
well constrained for all objects. None of the five objects has a clear characteristic of transition 
disks with large (e.g., AU-scale) inner holes \citep{williams2011}.

The flaring index $\beta$ and the scale height $h_{100}$ are key parameters for describing the disk geometry, 
which are thought to be different for disks around solar-type stars and their lower mass counterparts like brown dwarfs. 
For instance, \citet{szucs2010} investigated the Spitzer/IRAC and MIPS $24\,\mu{\rm{m}}$ photometries of ${\sim} 200$ 
stars in the Chamaeleon I star-forming region and found that disks around lower mass stars (spectral type later than M4.75) are 
generally flatter than the case of higher mass stars (spectral type earlier than M4.5). Coherent multiwavelength 
modeling suggests a typical flaring index $\beta \sim 1.25$ for T Tauri disks \citep{wolfp2003, walker2004, sauter2009, madlener2012, grafe2013}. 
\citet{harvey2012a} report PACS measurements of about 50 very low-mass stars and brown dwarfs with spectral types ranging from 
M3 to M9 in nearby regions. Through detailed modeling effort, they obtained a typically small (1.05-1.2) flaring index, which is 
consistent with our results for objects with spectral types similar to their sample. Theoretical models predict that disks around 
cooler stars should be more extended in the vertical direction \citep{walker2004}. However, our findings, together with previous 
studies \citep[e.g.,][]{harvey2012a, alves2013, olofsson2013}, show that both the T Tauri disks and brown dwarf disks feature similar 
scale heights, i.e., $5-20\,\rm{AU}$. The small flaring index ($\beta_{\rm{best-fit}}=1.01$) and scale height 
($h_{\rm{100,best-fit}}{=}2\,\rm{AU}$) indicate a flat and thin disk around TWA 30A, consistent with the effect 
of dust settling in disks. The disk of TWA 30B also shows signs of evolution, i.e., a small amount of flaring ($\beta_{\rm{best-fit}}=1.02$). 
Given the relatively old age of the TWA, our interpretations on the disk structure of TWA 30A and TWA 30B are consistent 
with theories of circumstellar disk evolution.   

\begin{figure}[!t]
\includegraphics[width=0.5\textwidth]{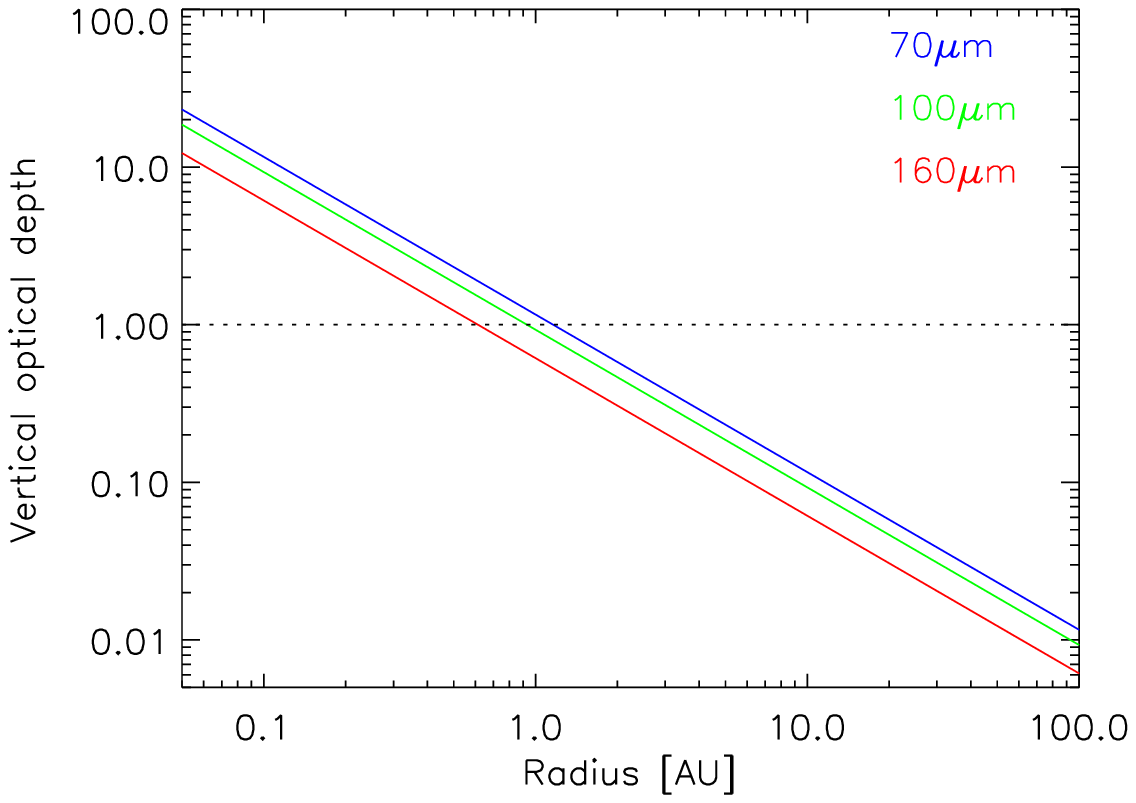}
   \caption{Optical depth in the vertical direction as a function of radius at {\it Herschel} far-IR wavelengths 
            for a typical brown dwarf disk model with ${\rm log_{10}}\,M_{\rm{disk}}/M_{\odot}=-4.5$ and dust properties 
            including grain size distribution identical to those used in the SED simulation, see Sect. \ref{sec:modeling}.}
\label{fig:tau}
\end{figure}

Photometry in the (sub-)mm is the best tool for studying the disk mass since the emission in this wavelength regime is 
almost optically thin and in the clear Rayleigh-Jeans part of the SED. However, the very faint (sub-)mm fluxes of brown 
dwarf disks (or disks around very low-mass stars) make their characterization extremely challenging, leading to a very 
small number of such detections to date \citep{scholz2006, ricci2012, ricci2013, mohantyg2013}. As shown 
by \citet{harvey2012a}, {\it Herschel} far-IR observations provide an alternative way 
to estimate the masses of faint disks as in our case.

 The five targets investigated here are detected with PACS at 
all wavelengths ($70\,\mu{\rm{m}}$, $100\,\mu{\rm{m}}$, and $160\,\mu{\rm{m}}$), except for TWA 30A
and TWA 31 at $160\,\mu{\rm{m}}$. Moreover, the observations taken with SCUBA-2 can place stringent upper limits on the disk 
masses for TWA 30A and TWA 30B. The Bayesian probability distributions of $M_{\rm disk}$ exhibit an obvious peak, 
with the most probable disk masses close to the best-fit values for all sources, indicating strong constraints 
on this parameter. The validity range of $M_{\rm disk}$ ($10^{-6}-10^{-4}\,M_{\odot}$) is far below the typical masses 
(i.e., $10^{-3}-10^{-1}\,M_{\odot}$) of disks around T Tauri stars located in various regions with ages ranging 
from ${\sim}1-2\,\rm{Myr}$ \citep[e.g., Taurus and Ophiuchus,][]{andrewsw2005, andrews2007, andrews2009},
through ${\sim}2-3\,\rm{Myr}$ \citep[e.g., IC348,][]{lee2011}, and to ${\sim}5-11\,\rm{Myr}$ 
\citep[e.g., TWA and Upper Scorpius,][]{mohantyg2013,mathews2012,carpenter2014}.

The lower disk masses around very low-mass stars and brown dwarfs are consistent with previous results from case studies and 
surveys \citep[e.g.,][]{broekhoven2014, harvey2012a}. Nevertheless, we should re-examine this conclusion because the determination of disk 
mass with far-IR photometry may be considered as lower limits owing to optical depth effects. As an illustration, Fig. \ref{fig:tau} 
shows the optical depth perpendicular to the disk as a function of radius, indicating that the inner region ($<$ a few AU) of 
typical brown dwarf disks are optically thick at {\it Herschel} wavelengths. The large portion of optically thin far-IR emission 
demonstrates that PACS bands are long enough to provide reasonable disk mass estimation of brown dwarfs.
The model presented in Fig. \ref{fig:tau} should be considered as the ``least favorable'' example since the 
assumed disk mass is higher than any of the best-fit results (see Table \ref{tab:bestfit}). Therefore, the best-fit models are 
in principle less optically thick in the vertical direction than the case shown in Fig. \ref{fig:tau}. 
Future ALMA observations in the (sub-)mm regime will enable more accurate masses to be determined for these (very) low-mass disks.

\begin{figure}[!t]
\includegraphics[width=0.5\textwidth]{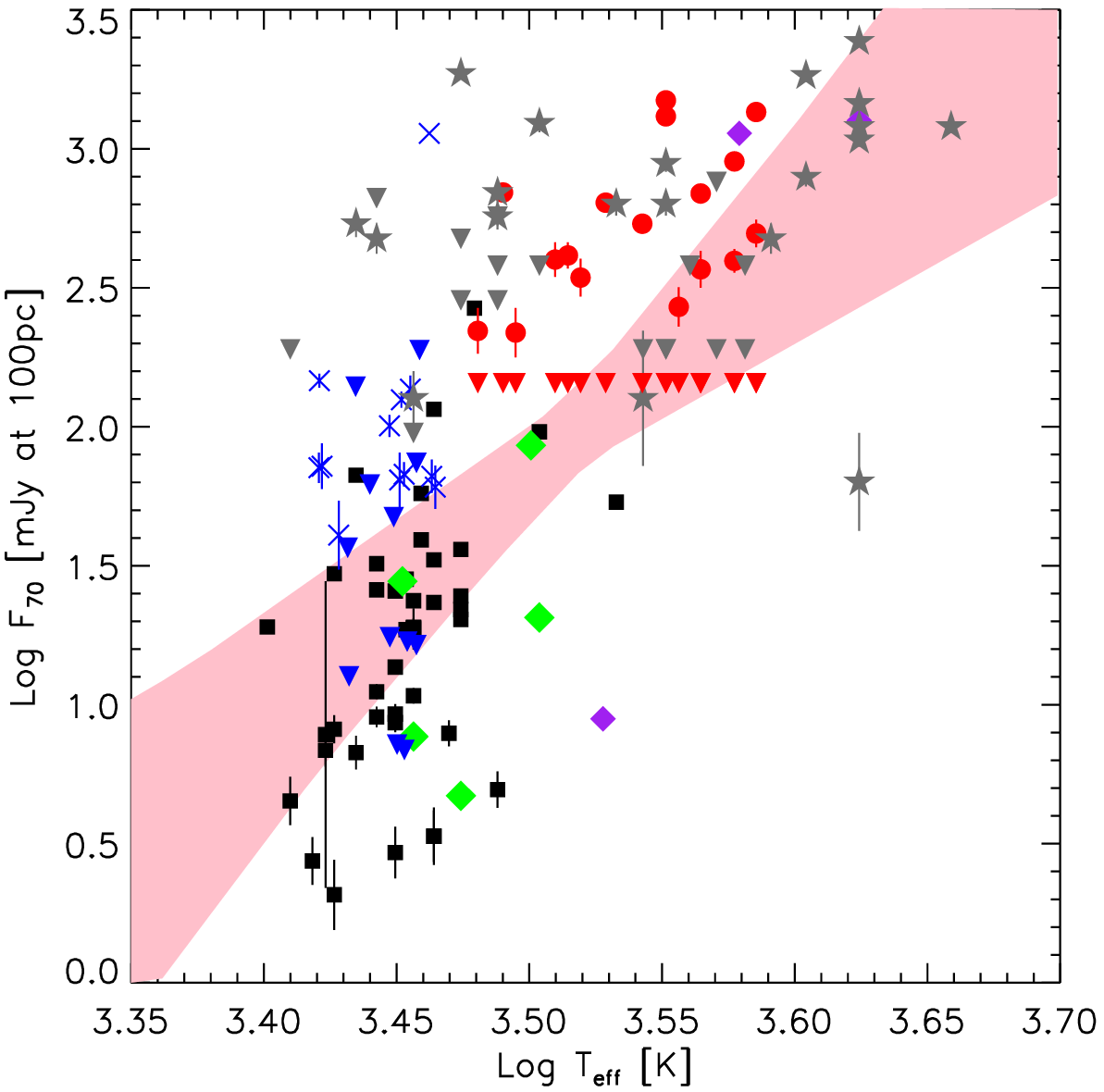}
   \caption{Fluxes at {\it Herschel} $70\,\mu{\rm{m}}$ ($F_{70}$) as a function of the stellar temperature ($T_{\rm{eff}}$). 
    Data points are collected from \citet{harvey2012a, harvey2012b} for a mix of nearby regions (black squares), \citet{olofsson2013} 
    for Chameleon I (red circles), \citet{spezzi2013} for Chameleon II (gray stars), \citet{alves2013} for $\rho$ Ophiuchi (blue crosses), 
    and from \citet{riviere-marichalar2013} (purple diamonds) and our program (green diamonds) for the TWA. Upper limits are indicated as upside down 
    triangles with the same color scheme. The $100\,\mu{\rm{m}}$ fluxes reported by \citet{olofsson2013} were converted to $70\,\mu{\rm{m}}$ 
    fluxes by multiplying them with a factor of 1.12, the median from very low-mass stars with detections at both wavelengths.
    The shaded region marks 95\% confidence intervals on the  ${\rm log} T_{\bf eff}\,{\sim}\,{\rm log} F_{70}$ relation, 
    derived from a Bayesian linear regression analysis that takes the errors and upper limits of measurements 
    into account \citep{kelly2007}.}
\label{fig:lumplots}
\end{figure}

\subsection{Dependence on the stellar properties}

Many observations have demonstrated that disk properties, such as the accretion rate and dust processing speed,  
depend on their host stellar properties like the effective temperature ($T_{\rm{eff}}$) and mass \citep{muzerolle2005, herczeg2009, pascucci2009, riaz2009}. 
Correlation analysis is therefore very common in disk studies, although the derivation of both disk and stellar properties always 
depends on model assumptions. The $M_{\rm disk}-M_{\star}$ correlation study is of particular interest because any correlation 
between these two quantities has important implications for planet formation theories.

\citet{andrewsr2013} analyzed for the first time mm continuum photometry of a sample of ${\sim}200$ Class II disks 
that is statistically complete for stellar hosts with spectral types earlier than M8.5 in the Taurus molecular cloud. 
They find a strong correlation between the mm luminosity (a good proxy for disk mass) and the spectral type of the 
host star, as well as the stellar mass. \citet{olofsson2013} derived the disk mass of 17 low-mass M-type stars from 
modeling their broadband SEDs, including {\it Herschel} far-IR photometry. However, they did not find a clear trend 
among the $M_{\rm disk}-M_{\star}$ diagram, probably because of the detection biases and the optical depth effect in the 
far-IR. The disk mass is not a direct observable, and its estimation needs additional assumptions like the dust 
opacity. Therefore, it is not appropriate to analyze the modeling results collected from the literature 
that make different assumptions.

\begin{figure}[!t]
\includegraphics[width=0.5\textwidth]{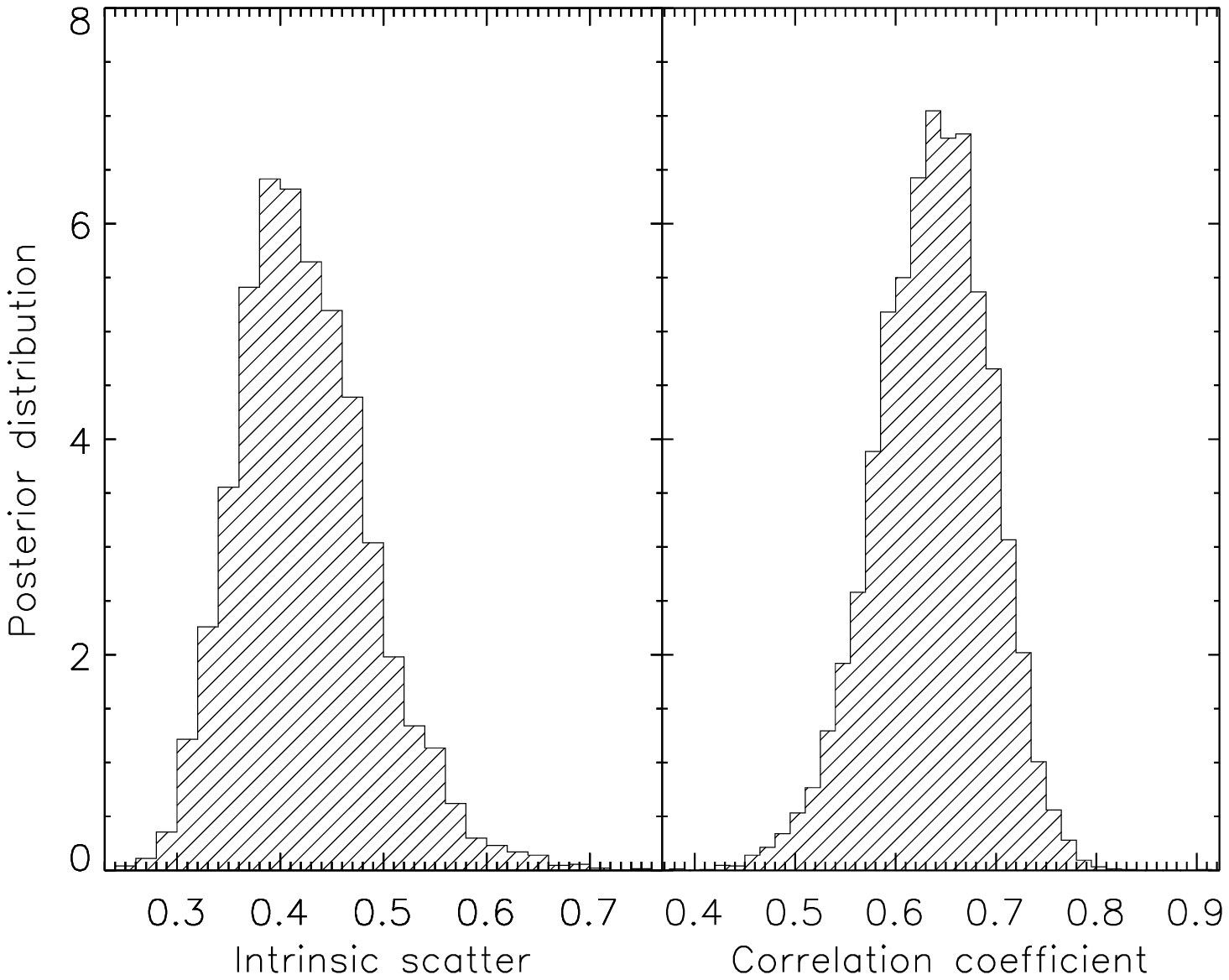}
\caption{Marginal posterior distributions of the intrinsic dispersion around the regression line and linear correlation coefficient.}
\label{fig:corr}
\end{figure}

Considering the factors mentioned above, we instead compared the flux ($F_{70}$) at the {\it Herschel} $70\,\mu{\rm{m}}$ with 
the effective temperature of the host stars. In Fig. \ref{fig:lumplots}, we show the results from 
our program and from the literature \citep{harvey2012a, harvey2012b, olofsson2013, spezzi2013, alves2013, riviere-marichalar2013}. 
All fluxes are scaled assuming a distance of $100\,\rm{pc}$. The conversion from spectral type to temperature from \citet{herczeg2014} 
was used to calculate $T_{\rm{eff}}$ in a consistent manner. The stellar effective temperature and $70\,\mu{\rm{m}}$ flux are 
correlated, though with an especially large dispersion at low temperatures. We used Spearman's rank correlation coefficient ($\rho$) 
and its two-sided probability ($P$) to test the correlation \citep{kim2009}. The resulting $\rho=0.685$ and $p\ll0.05$ suggest a 
statistically significant correlation between the amount of {\it Herschel} far-IR emission and the stellar temperature. 
This conclusion is also supported by Pearson's linear correlation coefficient of $\gamma=0.728$, a value close to 1. 

To take the measurement errors and upper limits into account, we used the Bayesian linear regression method developed 
by \citet{kelly2007} to quantify the correlation, assuming an intrinsic linear dependence between ${\rm log} T_{\rm{eff}}$ and ${\rm log} F_{70}$. 
The errors of $T_{\rm{eff}}$ are assumed to be $100\,\rm{K}$ for all targets in this analysis. The results are presented in Fig. \ref{fig:corr}. 
The histogram of the associated correlation coefficient in this figure confirms the tight relationship determined from the correlation 
tests. The lefthand panel of Fig. \ref{fig:corr} shows the variance in the residuals between the estimated regression line and the
dependent variable $F_{70}$, reflecting typical dispersion in $F_{70}$ around the regression line at any given $T_{\rm{eff}}$.
The 95\% pointwise confidence intervals on the regression line are indicated as shaded regions in Fig. \ref{fig:lumplots}. 
As {\it Herschel} $160\,\mu{\rm{m}}$ images of faint disks are easily contaminated with the extended emission from the background, sources 
with solid detections at $160\,\mu{\rm{m}}$ so far are not sufficient in number for this correlation analysis.

The effective temperature is a good proxy for the stellar mass ($M_{\star}$) of low-mass stars and brown dwarfs, because 
the pre-main-sequence tracks of low-mass stars and brown dwarfs at ages $1{\sim}10\,\rm{Myr}$ are perpendicular to the 
temperature axis, i.e., without any obvious turning tendency. As shown in Fig. \ref{fig:tau}, PACS $70\,\mu{\rm{m}}$ emission 
approximately probe the brown dwarf disk mass, affected mainly by the optically thick emission from the disk inner region 
and limited by the fact that $70\,\mu{\rm{m}}$ is not completely into the Rayleigh-Jeans part of the SED. Given the high 
ratio of optically thin-to-thick regions of the disk, the uncertainty of disk mass estimated from far-IR photometry should not exceed 
one order of magnitude. This can be justified by the validity ranges of $M_{\rm{disk}}$ deduced from Bayesian analysis (see Table \ref{tab:bestfit}). 
Therefore, the $F_{70}\,{\sim}\,T_{\rm{eff}}$ correlation implies that the linear 
scaling $M_{\rm{disk}} \propto M_{\star}$ observed mainly in T Tauri and Herbig AeBe stars \citep{andrewsr2013} may extend 
down to the brown dwarf regime. This is in line with the reliable finding from (sub-)mm studies that the disks around brown dwarf 
hosts are intrinsically low-mass \citep[e.g.,][]{klein2003, scholz2006, ricci2013, broekhoven2014}. 

The $M_{\rm{disk}} \propto M_{\star}$ relation, if present, suggests that the disk-to-host mass ratio is basically independent 
of $M_{\star}$ for brown dwarfs, which agrees with the result found for T Tauri stars \citep{williams2011}. Our correlation study 
is based on a sample of targets that span a wide range of age (i.e., with different evolutionary stages), which may play a role of 
broadening the observed relation. Moreover, the detection bias can also affect the correlation since most other samples have worse 
sensitivity to disk mass compared to TWA sources because they are located at larger distances and observed with shorter exposure times. 
For example, the data points taken from \citet{spezzi2013} and \citet{olofsson2013} are based on the {\it Herschel} Gould Belt survey \citep{andre2010} 
that is not sensitive enough to detect very faint disks as compared to dedicated projects like our program and the one 
in \citet{harvey2012a}. The ambiguity of non-detections by current PACS observations may cause additional spread in $F_{70}$ and in 
turn weaken the $M_{\rm{disk}} {\sim} M_{\star}$ relation. Therefore, we emphasize that the correlation 
between $M_{\rm{disk}}$ and $M_{\star}$ speculated here is not robust because it relies not only on the completeness and depth 
of the far-IR survey but also on the accuracy of the determined disk mass, both of which cannot be directly included in the 
analysis especially given that the available data are indeed limited. Deeper far-IR and more sensitive (sub-)mm 
photometry survey are required to improve constraints on the underlying morphology of the relationship 
between $M_{\rm{disk}}$ and $M_{\star}$ in the low stellar mass regime.

\section{Summary}
We measured far-IR photometry for disks around five low-mass stars and
brown dwarfs in the TWA. By combining with previous measurements
at shorter wavelengths, these new datasets enable us to construct observed SEDs 
with extended coverage, providing a valuable opportunity to investigate the 
properties of disks in the TWA. We performed detailed SED analysis for our 
targets with the radiative transfer code \texttt{MC3D} and the hybrid fitting approach. 
Using the Bayesian inference method, we evaluated the constraints on disk 
parameters obtained from the modeling effort. 

Our results show that the disks around low-mass stars or brown dwarfs are generally
flatter than their higher mass counterparts, but the range of disk mass extends to 
well below the value found in T Tauri stars. The radiative transfer simulation 
demonstrates that typical brown dwarf disks are vertically optically thin in far-IR emission 
over a large portion of radial distances, which indicates that the 
disk mass of brown dwarfs can be roughly probed by far-IR observation. A comparison between 
the {\it Herschel} far-IR photometry and the stellar temperature displays a significant correlation 
between these two quantities, suggesting that the linear scaling $M_{\rm{disk}} \propto M_{\star}$ 
observed mainly in T Tauri and Herbig AeBe stars may exist down to the brown dwarf regime.
With unprecedented capacity, future ALMA observations will be able to extend the SEDs to the mm wavelengths, 
which will help us to improve the constraints on the disk mass and geometry and therefore to better 
understand the formation and evolution of very low-mass stars and brown dwarfs.

\acknowledgements
We thank the anonymous referee for valuable comments that improved the manuscript. 
We thank Hongchi Wang and Zhibo Jiang for useful discussions.
Y.L. acknowledges the support by the Natural Science Foundation of Jiangsu Province of China (Grant No. BK20141046).
G.J.H. is supported by the Youth Qianren Program  of the National Science Foundation of China.
This publication makes use of data products from the Wide-field Infrared Survey Explorer, which 
is a joint project of the University of California, Los Angeles, and the Jet
Propulsion Laboratory/California Institute of Technology, funded by the National Aeronautics 
and Space Administration. PACS has been developed by a consortium of institutes led by 
MPE (Germany) and including UVIE (Austria); KU Leuven, CSL, IMEC (Belgium); CEA, LAM (France); 
MPIA (Germany); INAF- IFSI/OAA/OAP/OAT, LENS, SISSA (Italy); IAC (Spain). This work is 
supported by the Strategic Priority Research Program ``The Emergence of Cosmological Structures'' 
of the Chinese Academy of Sciences, Grant No. XDB09000000.

\bibliographystyle{aa}
\bibliography{bdref}

\begin{appendix}

\section{Disk parameters for the model grids and Bayesian probability analysis}
\begin{table}[!h]
 \centering
   \caption{Disk parameters for the grids of models.}
    \begin{tabular}{ll}
     \hline
   Parameter  &   Values  \\
     \hline
   $R_{\rm in}$ [$R_{\rm{sub}}$]:  & 1, 2.13, 4.54, 9.68, 20.64, 44.01, 93.82, 200 \\ % new variable
   ${\rm log}_{10}(M_{\rm disk}/\rm M_{\odot})$: & $-$6, $-$5.5, $-$5, $-$4.5, $-$4, $-$3.5, $-$3, $-$2.5, $-$2    \\
   $\beta$: & 1.0, 1.05, 1.1, 1.15, 1.2, 1.25      \\
   $h_{\rm 100}$ [AU]:  & 2, 5, 8, 11, 14, 17, 20  \\
   $i$ [$^{\circ}$]:  & 15, 30, 45, 60, 75, 90    \\
   \hline \\
   \end{tabular}
\label{tab:sampling}
\end{table}

\begin{figure*}[!t]
\centering
\begin{minipage}[c]{0.4\textwidth}
\centering
\includegraphics[width=2.1in]{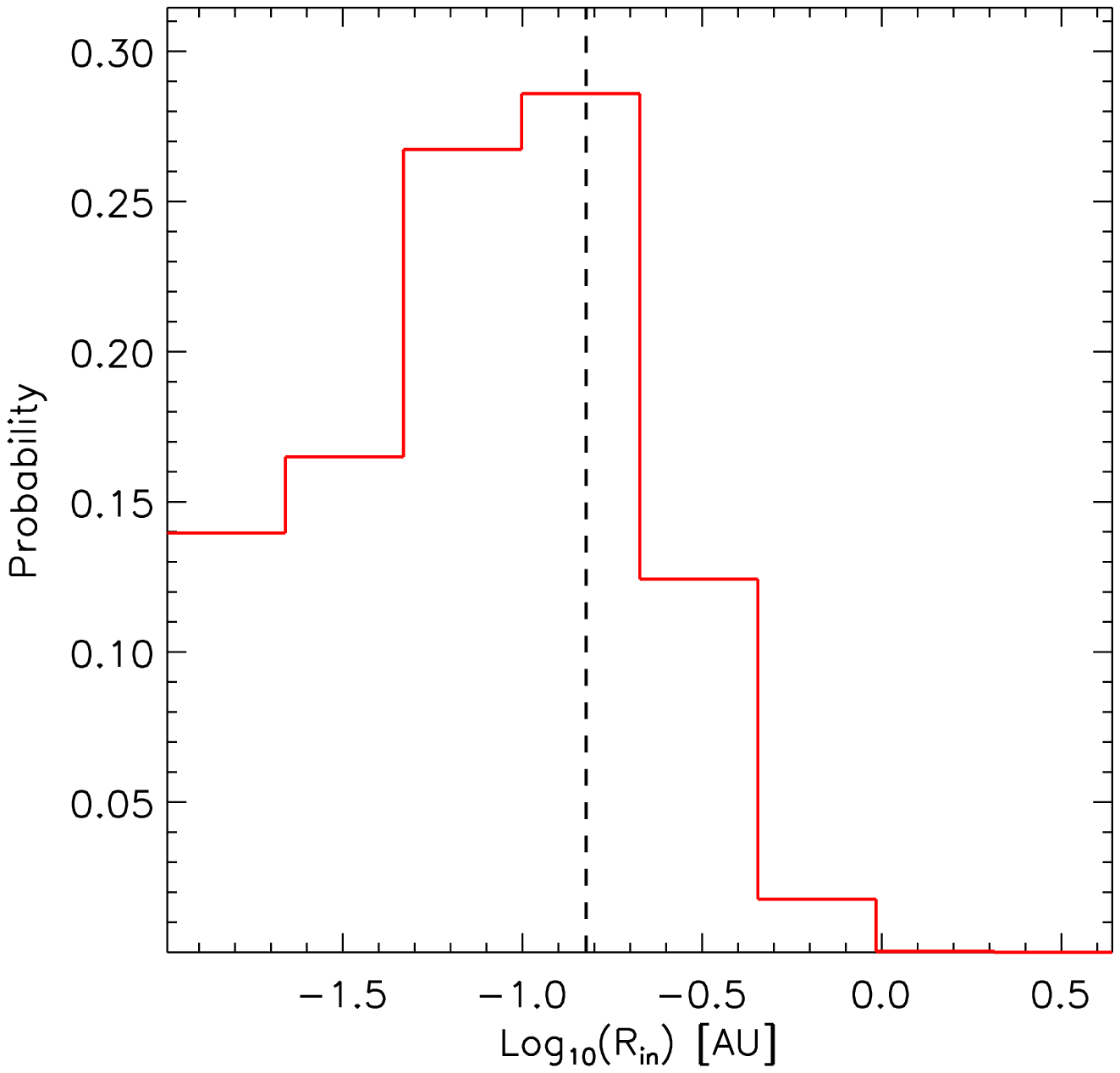}
\includegraphics[width=2.1in]{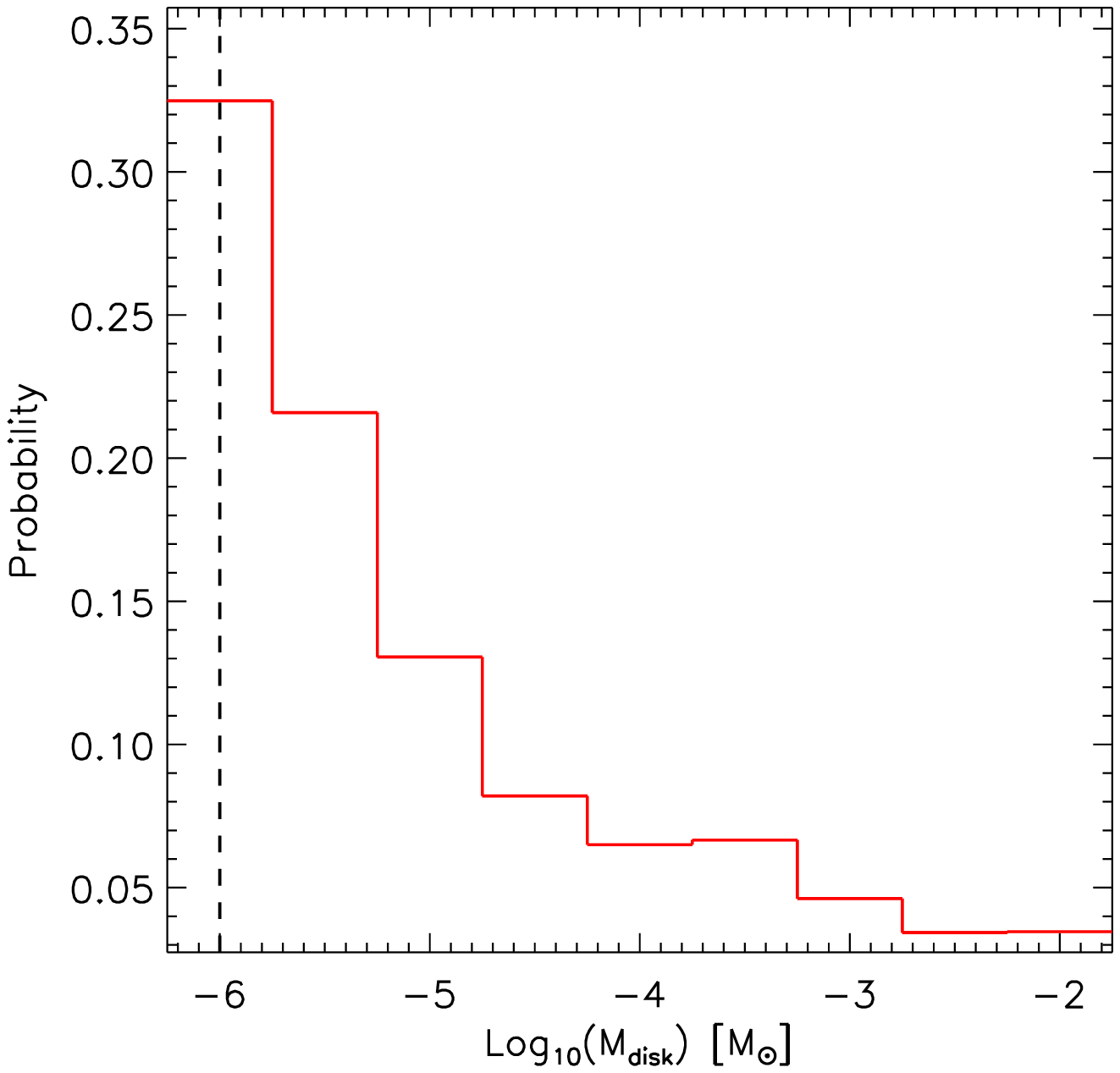}
\end{minipage}
\begin{minipage}[c]{0.4\textwidth}
\centering
\includegraphics[width=2.1in]{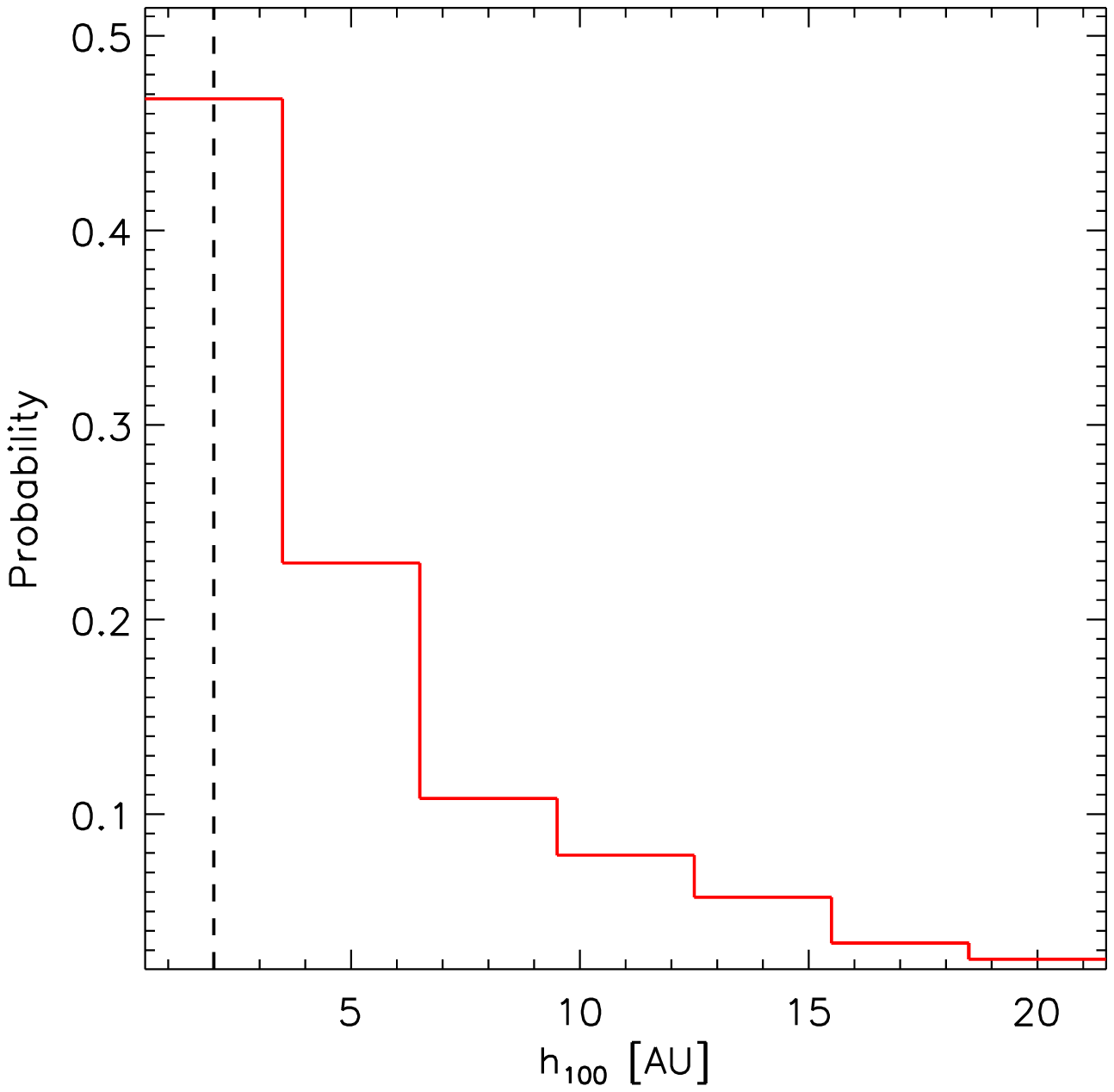}
\includegraphics[width=2.1in]{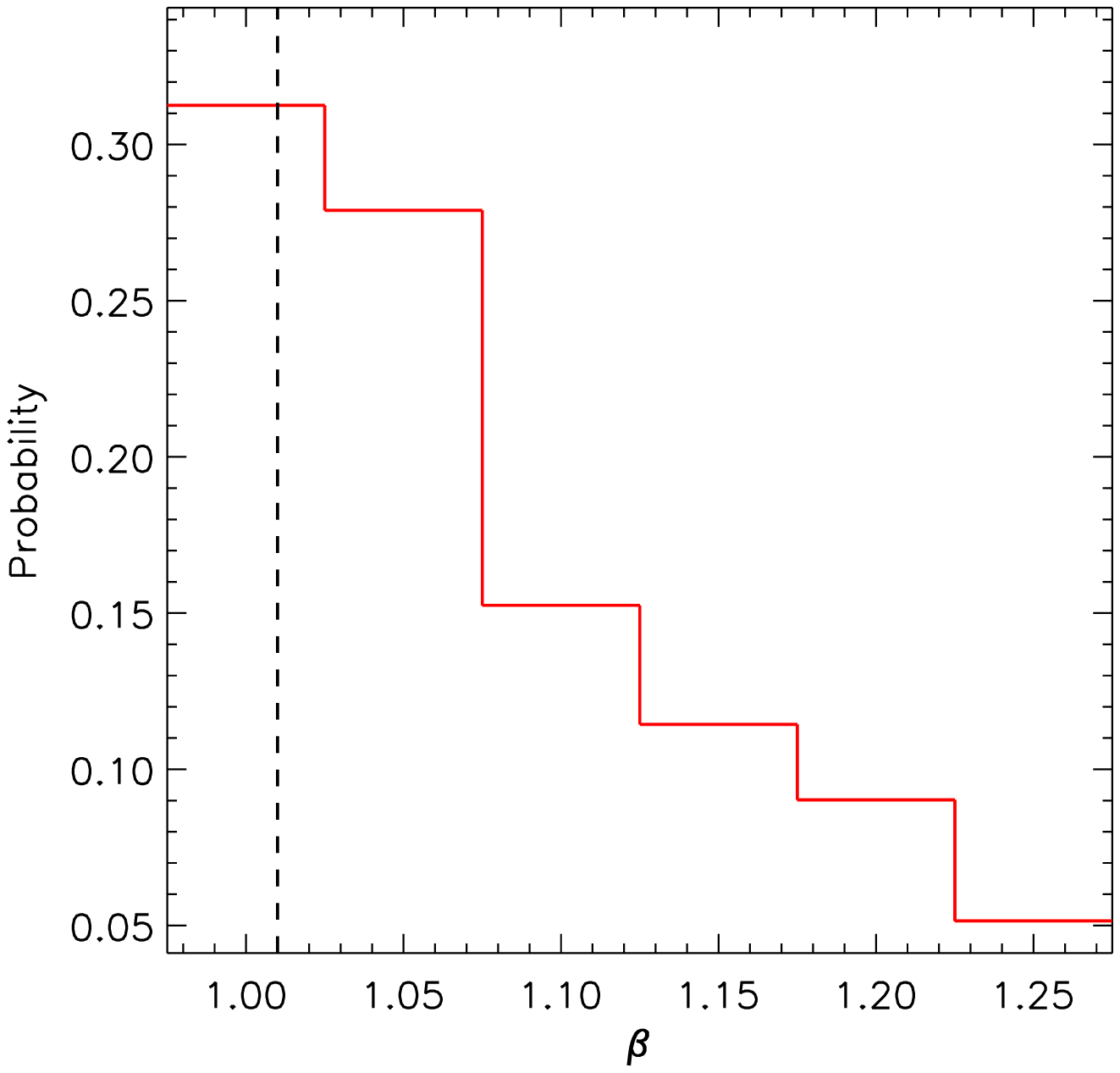}
\end{minipage}
\begin{minipage}[c]{0.4\textwidth}
\centering
\includegraphics[width=2.1in]{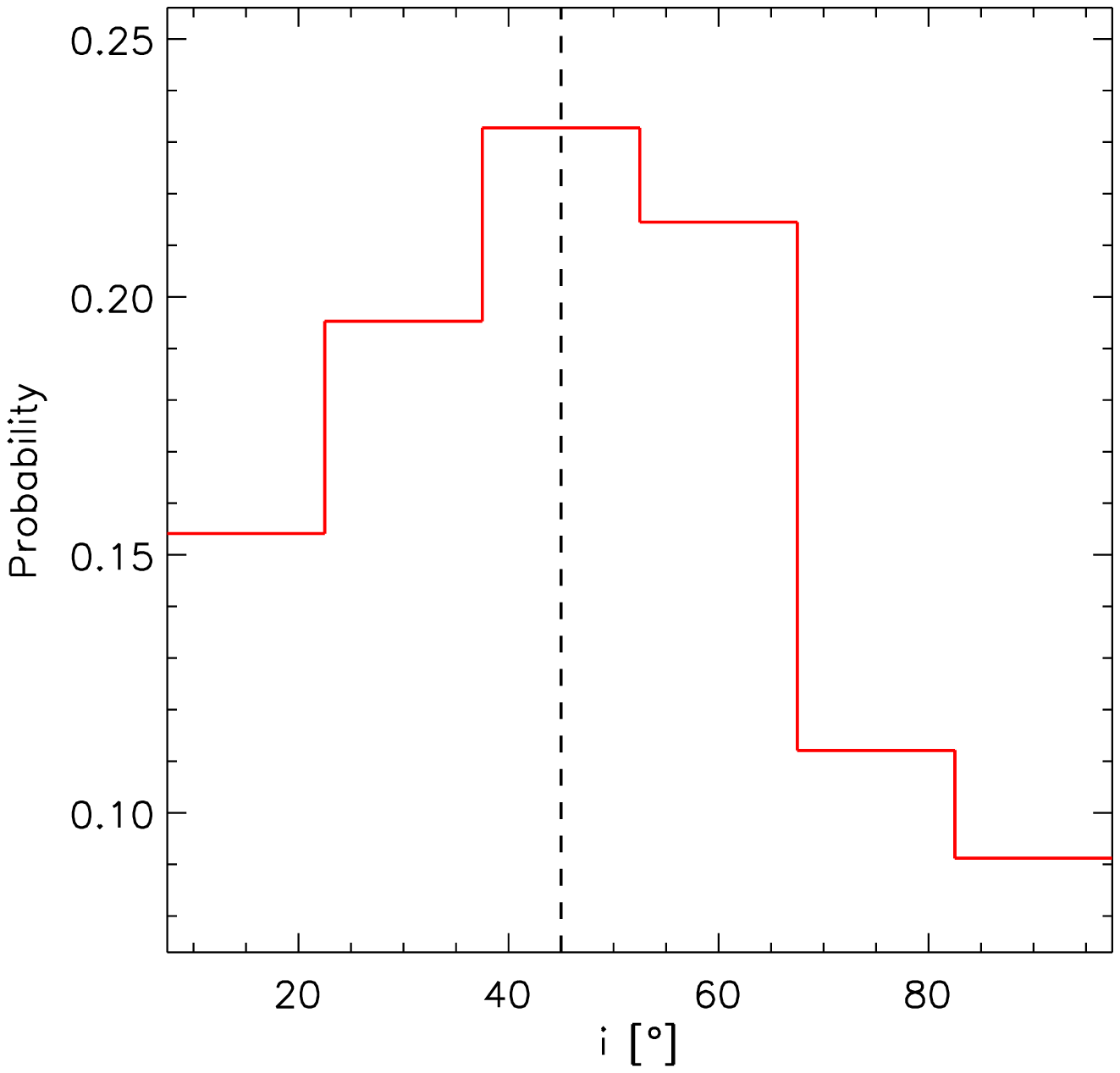}
\end{minipage}
\caption{Probability distributions of model parameters for TWA 30A. The vertical dashed lines denote the best-fit values.}
\end{figure*}

\newpage
\begin{figure*}[t]
\centering
\begin{minipage}[c]{0.4\textwidth}
\centering
\includegraphics[width=2.1in]{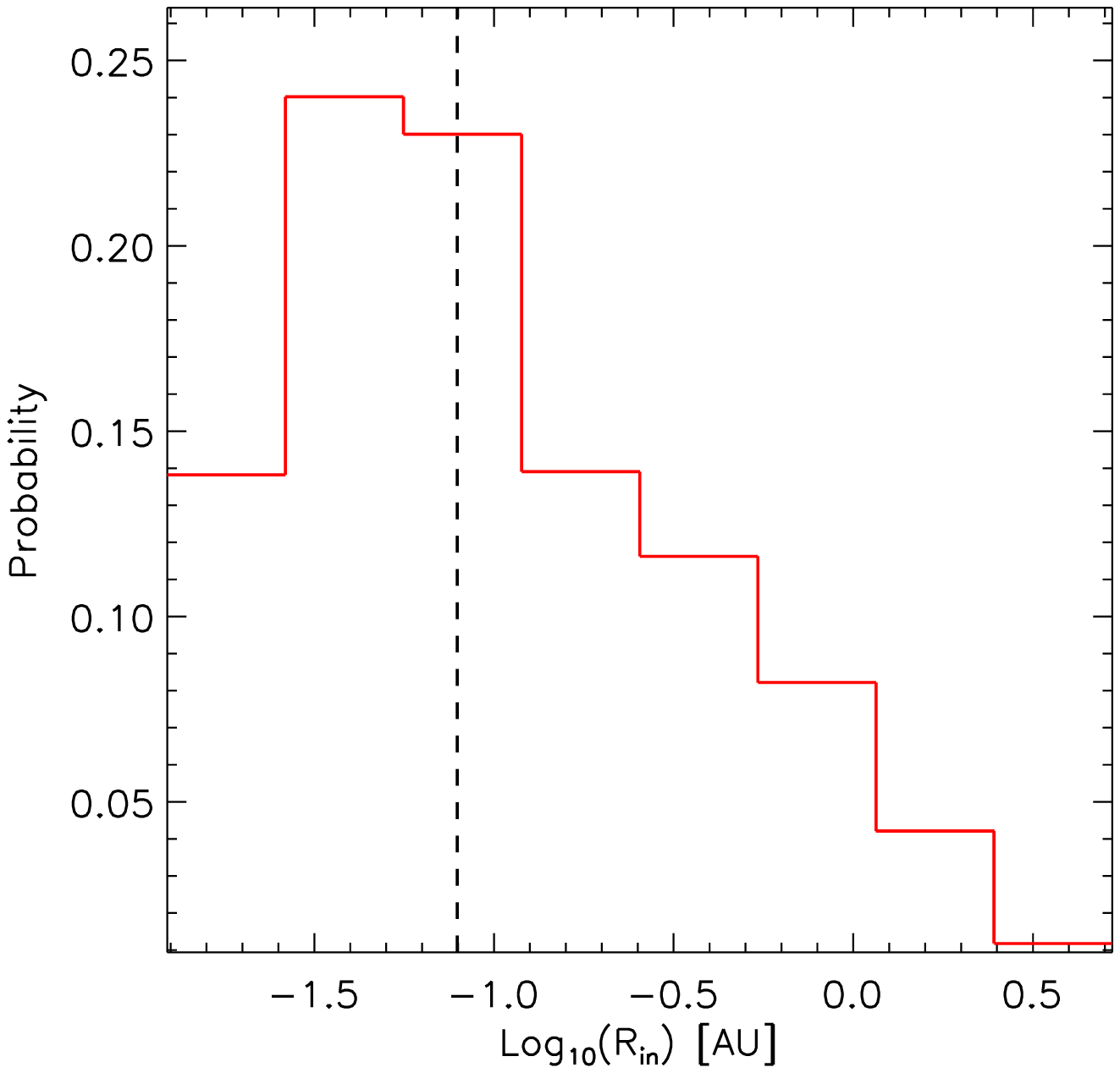}
\includegraphics[width=2.1in]{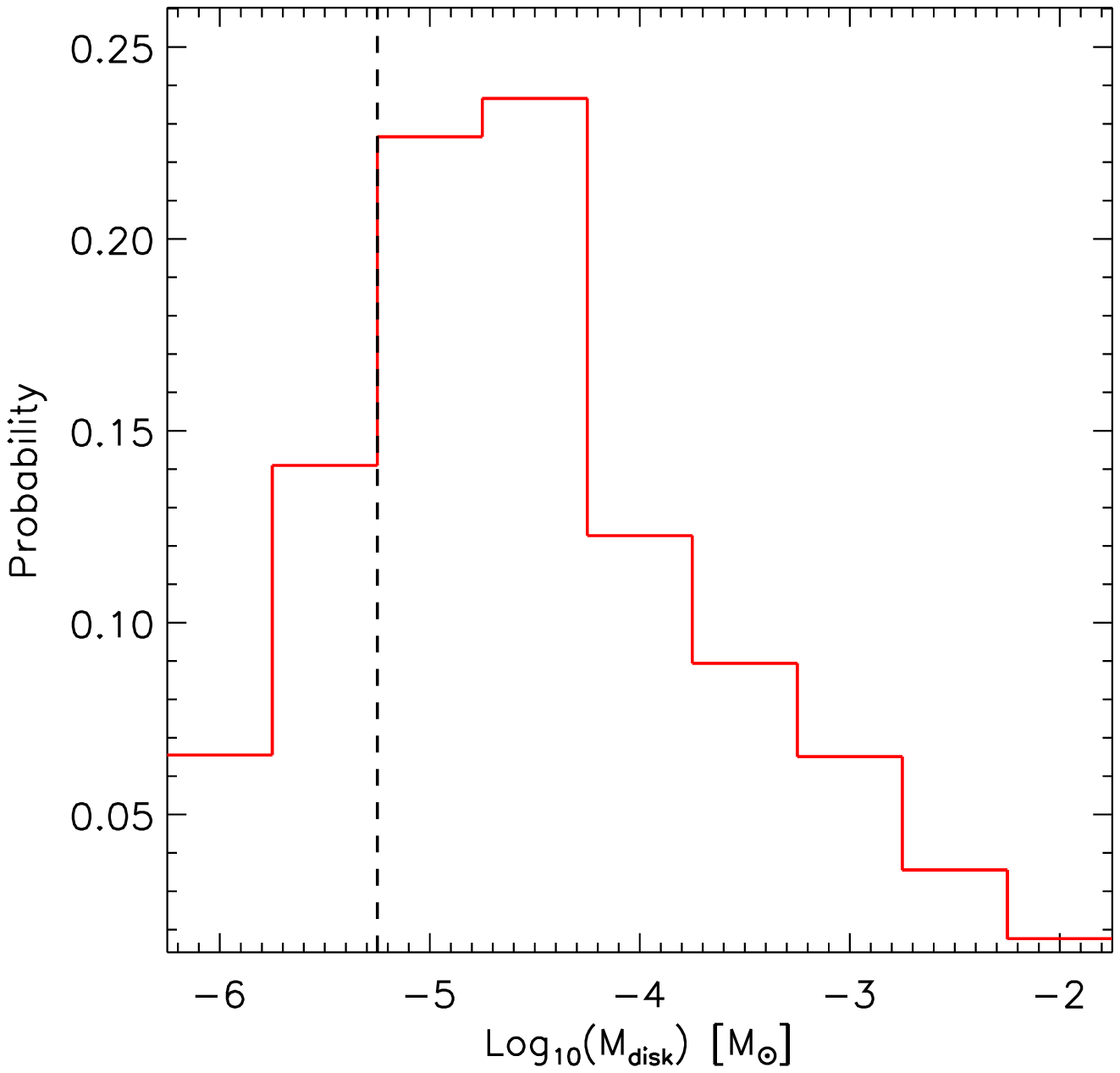}
\end{minipage}
\begin{minipage}[c]{0.4\textwidth}
\centering
\includegraphics[width=2.1in]{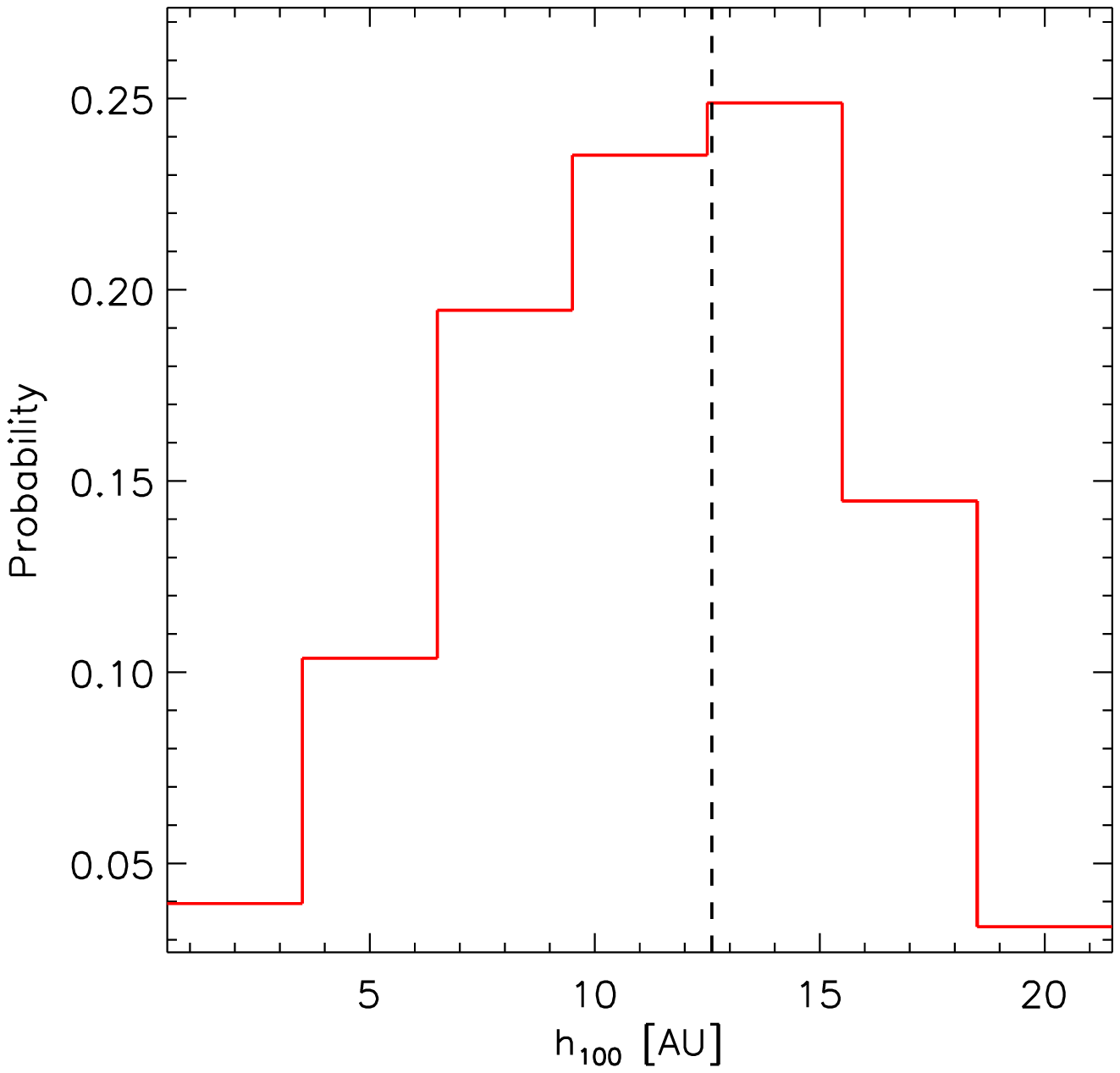}
\includegraphics[width=2.1in]{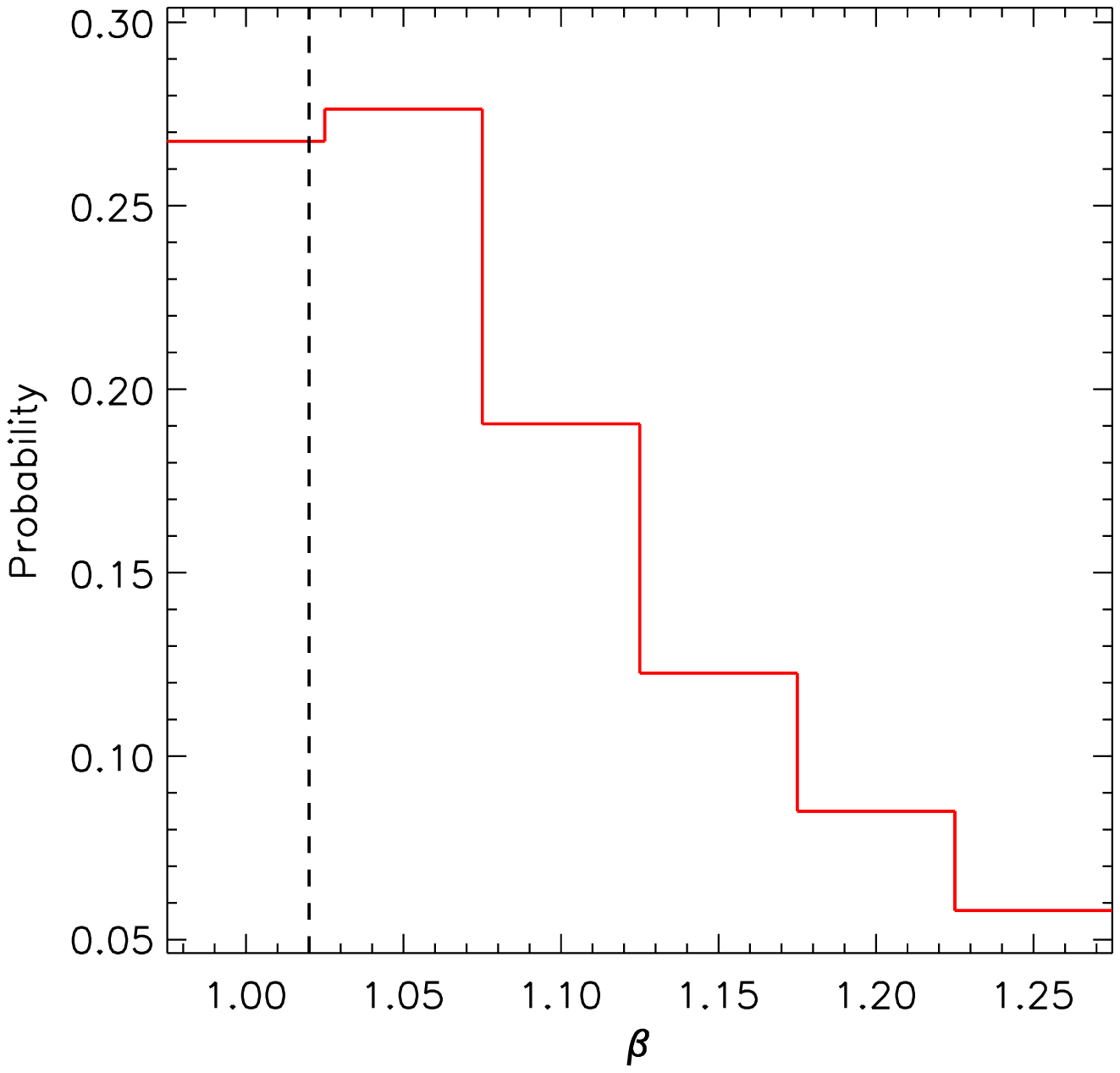}
\end{minipage}
\begin{minipage}[c]{0.4\textwidth}
\centering
\includegraphics[width=2.1in]{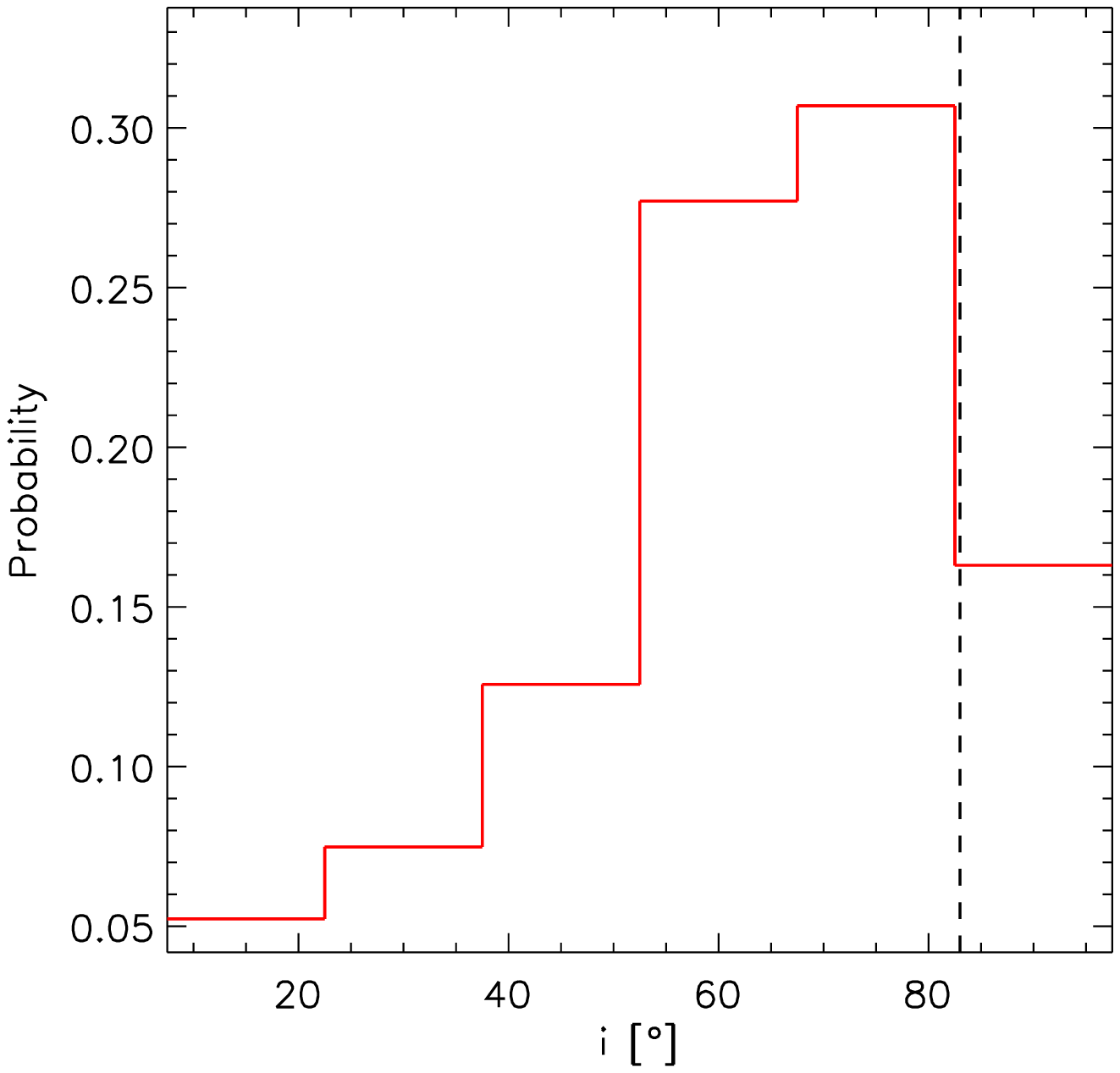}
\end{minipage}
\caption{Probability distributions of model parameters for TWA 30B. The vertical dashed lines denote the best-fit values.}
\label{fig:twa30b}
\end{figure*}

\newpage
\begin{figure*}[t]
\centering
\begin{minipage}[c]{0.4\textwidth}
\centering
\includegraphics[width=2.1in]{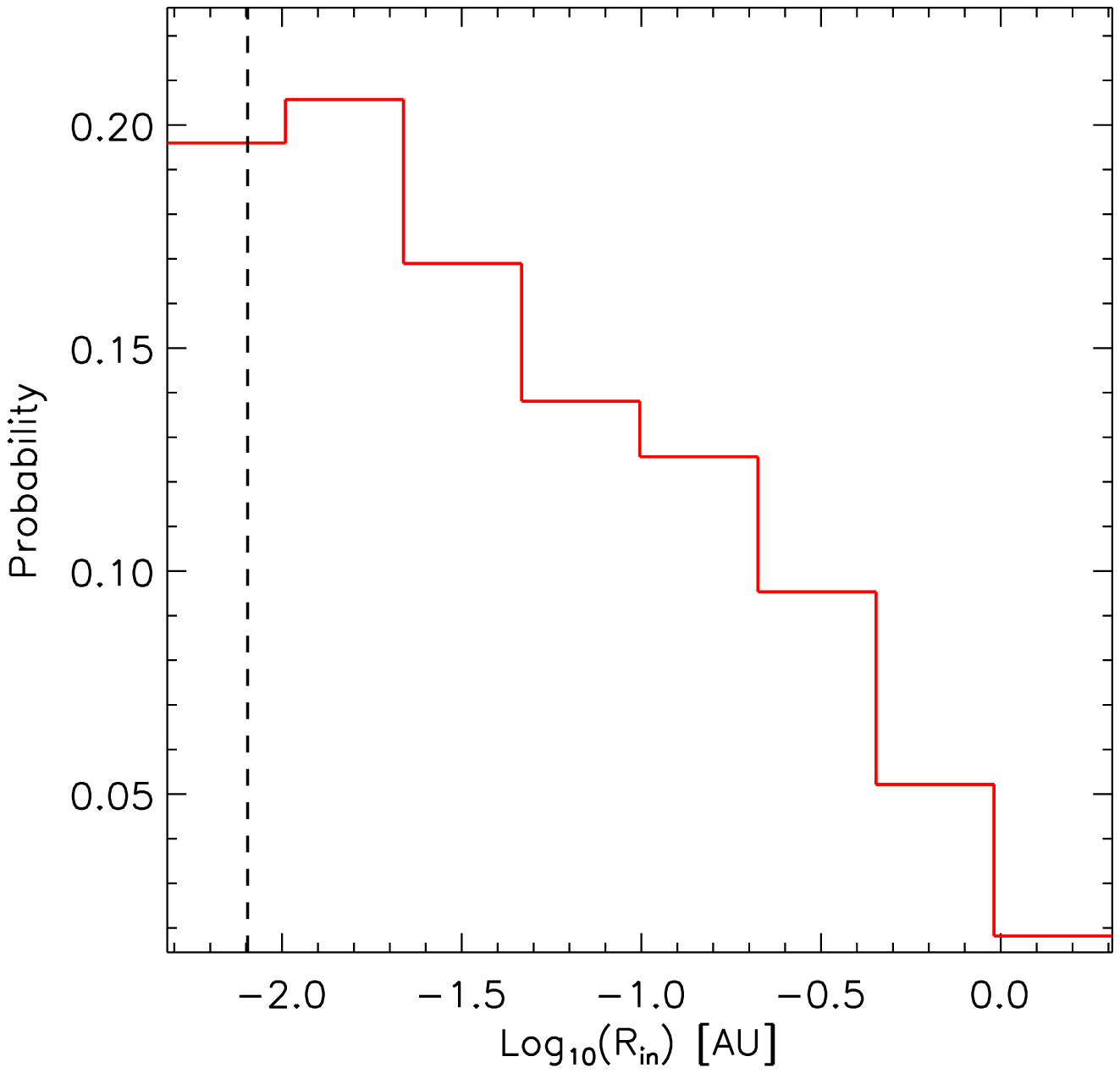}
\includegraphics[width=2.1in]{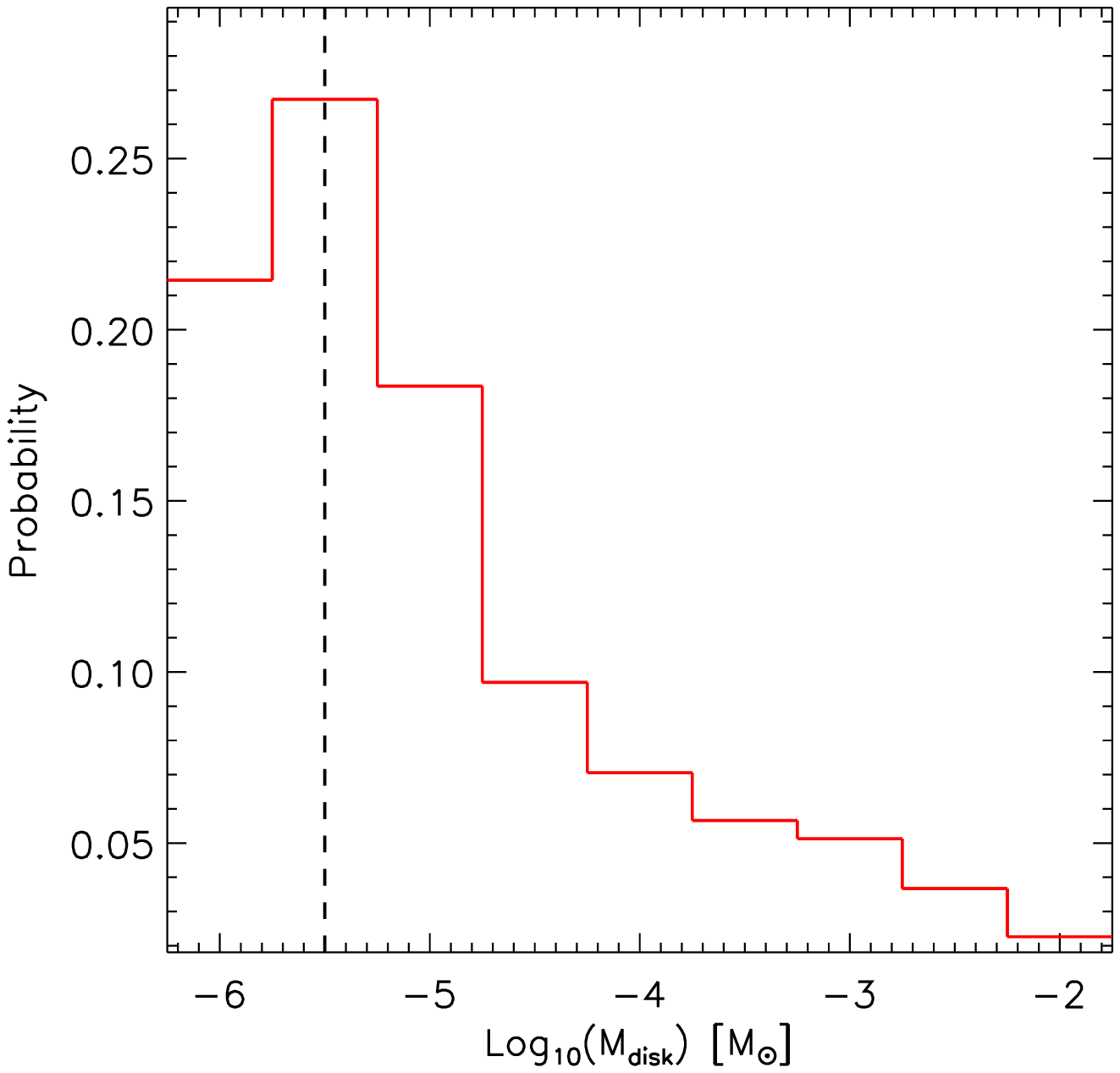}
\end{minipage}
\begin{minipage}[c]{0.4\textwidth}
\centering
\includegraphics[width=2.1in]{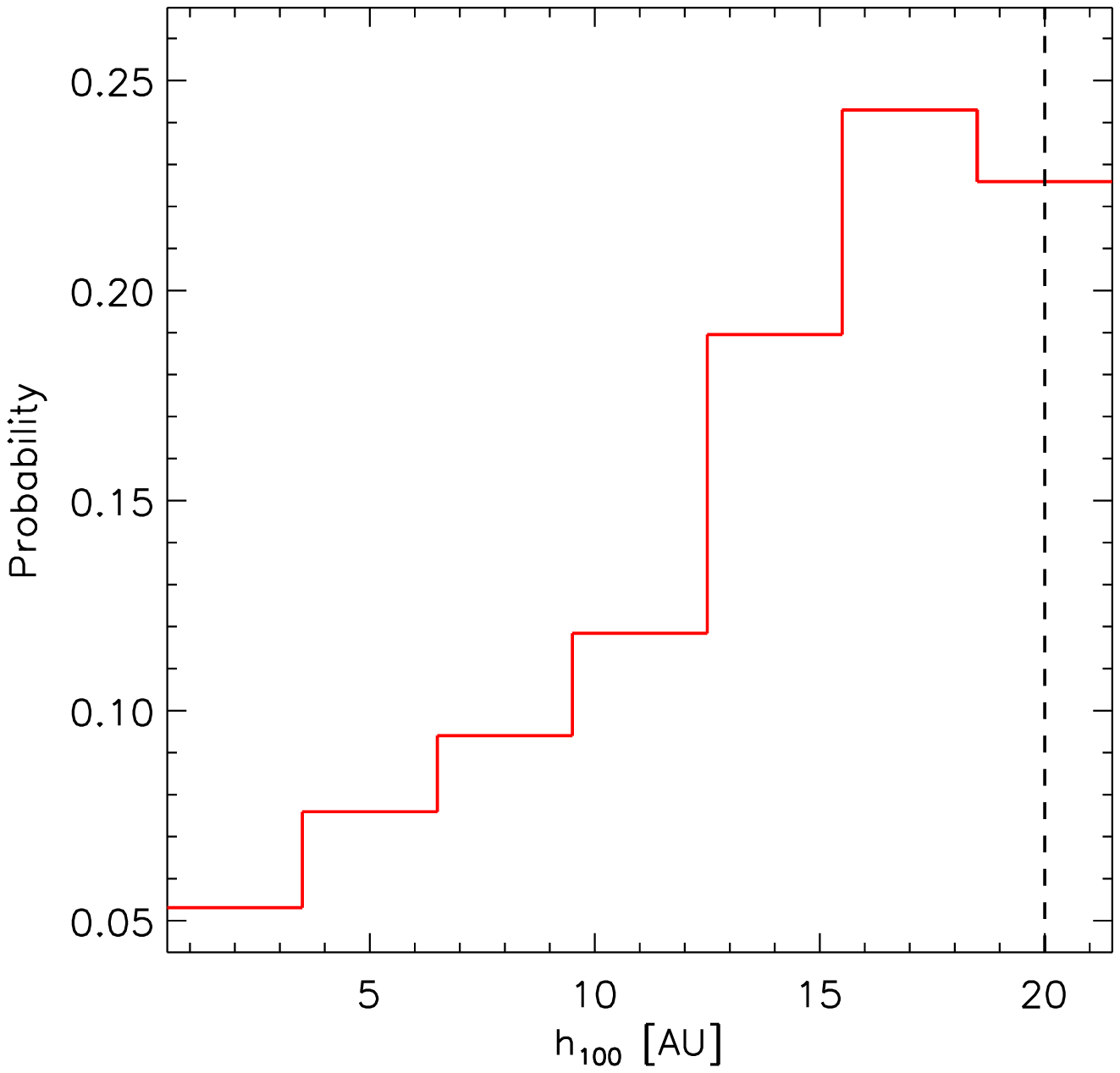}
\includegraphics[width=2.1in]{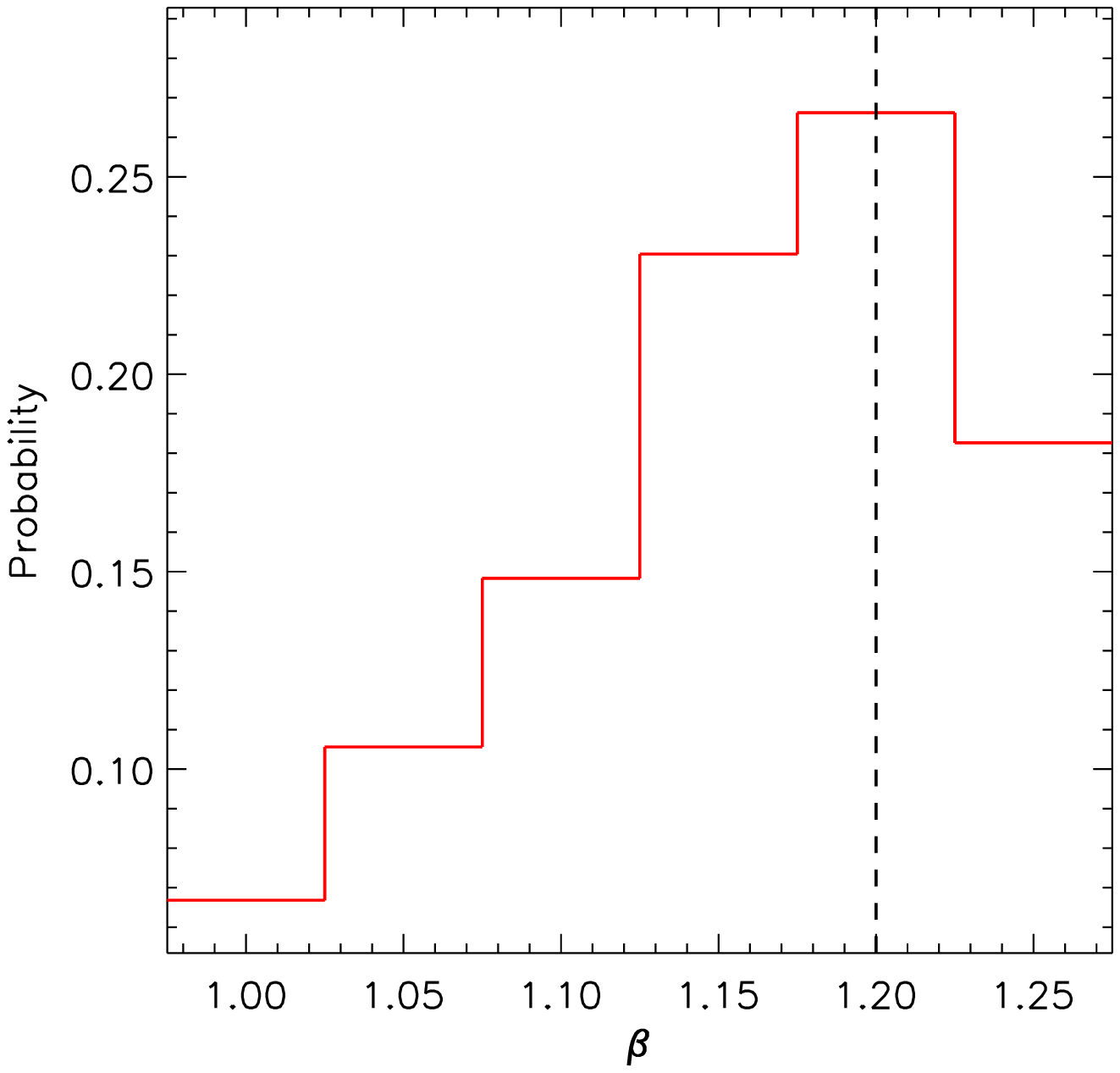}
\end{minipage}
\begin{minipage}[c]{0.4\textwidth}
\centering
\includegraphics[width=2.1in]{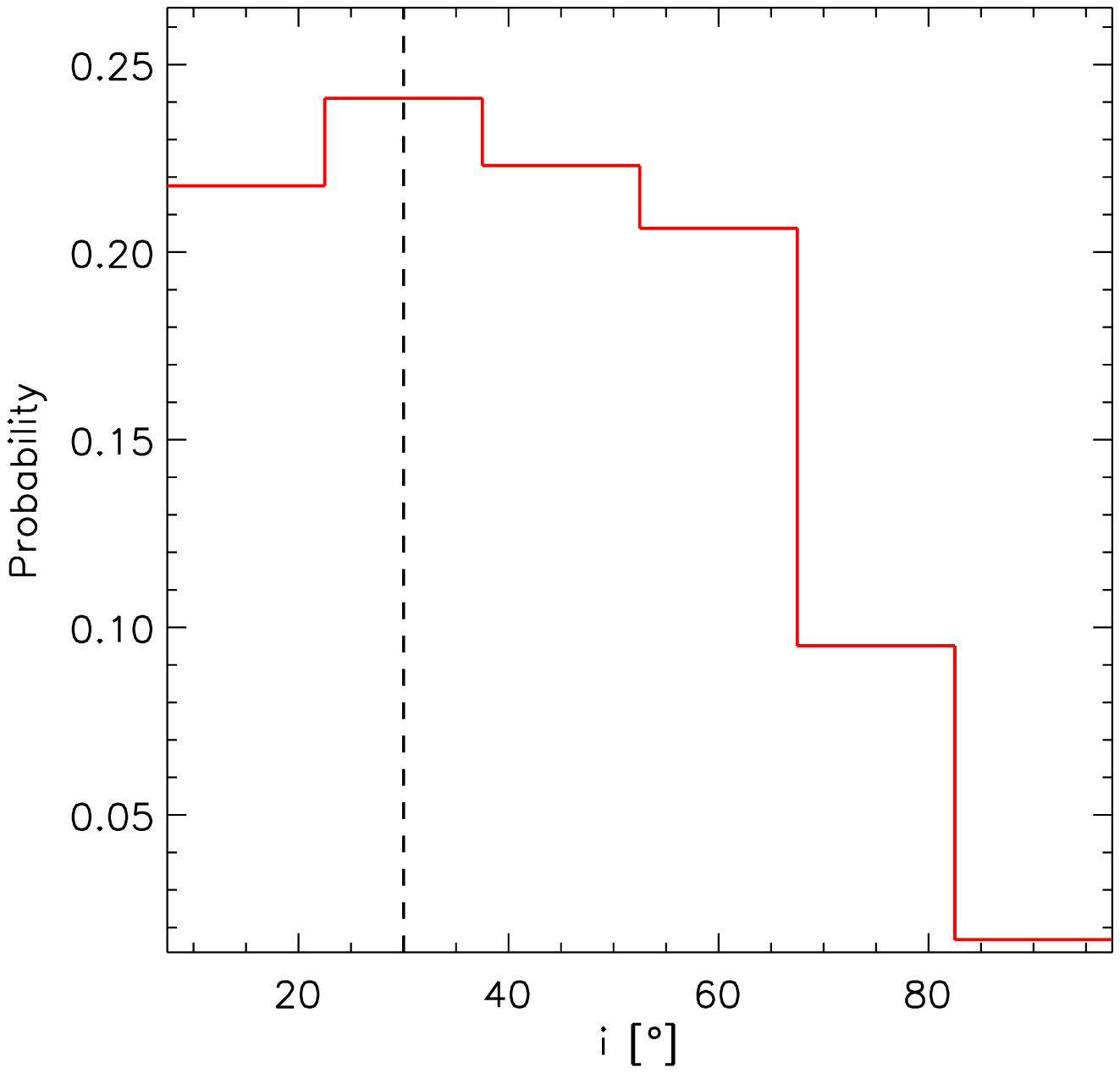}
\end{minipage}
\caption{Probability distributions of model parameters for TWA 31. The vertical dashed lines denote the best-fit values.}
\end{figure*}

\newpage
\begin{figure*}[t]
\centering
\begin{minipage}[c]{0.4\textwidth}
\centering
\includegraphics[width=2.1in]{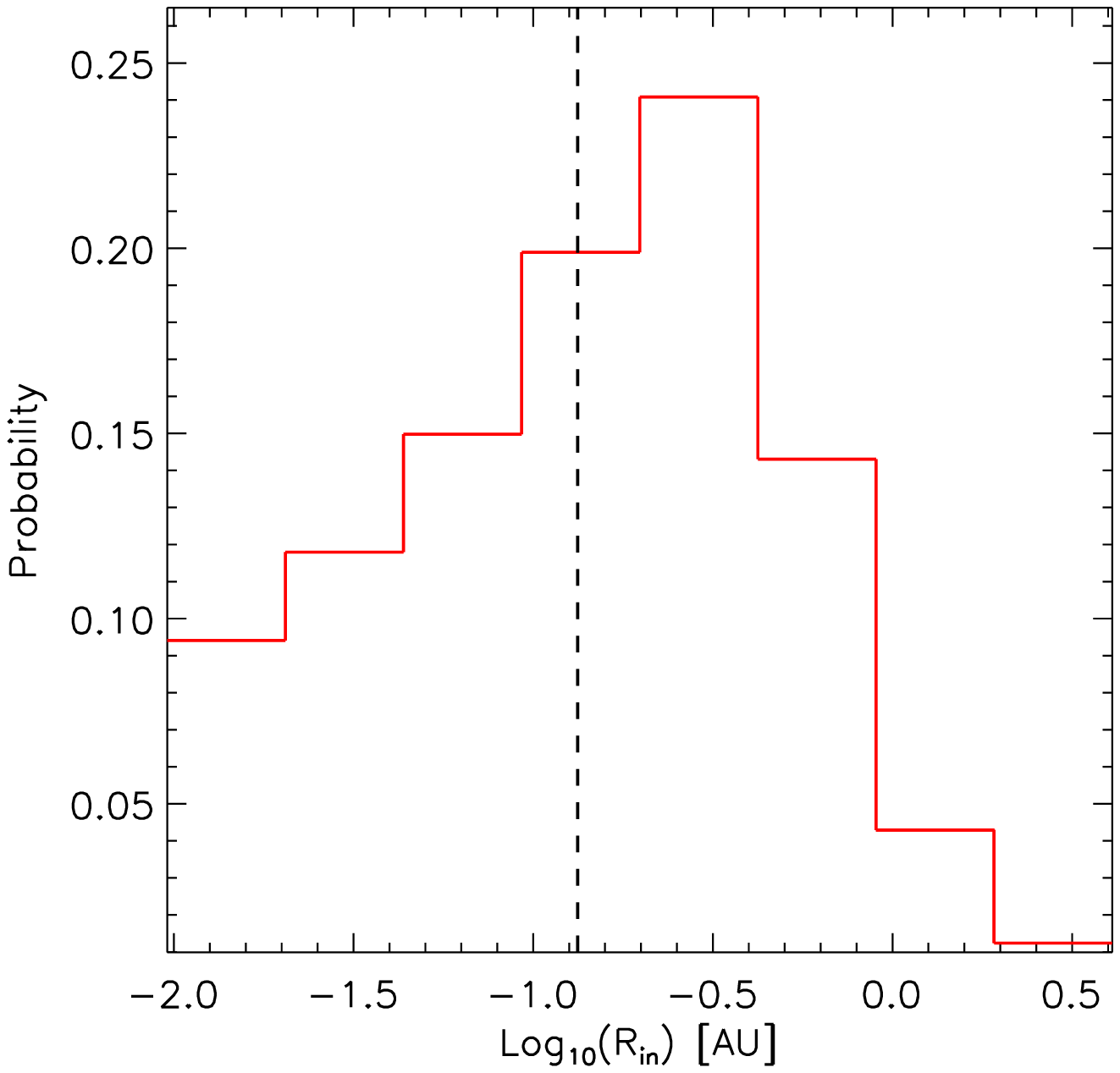}
\includegraphics[width=2.1in]{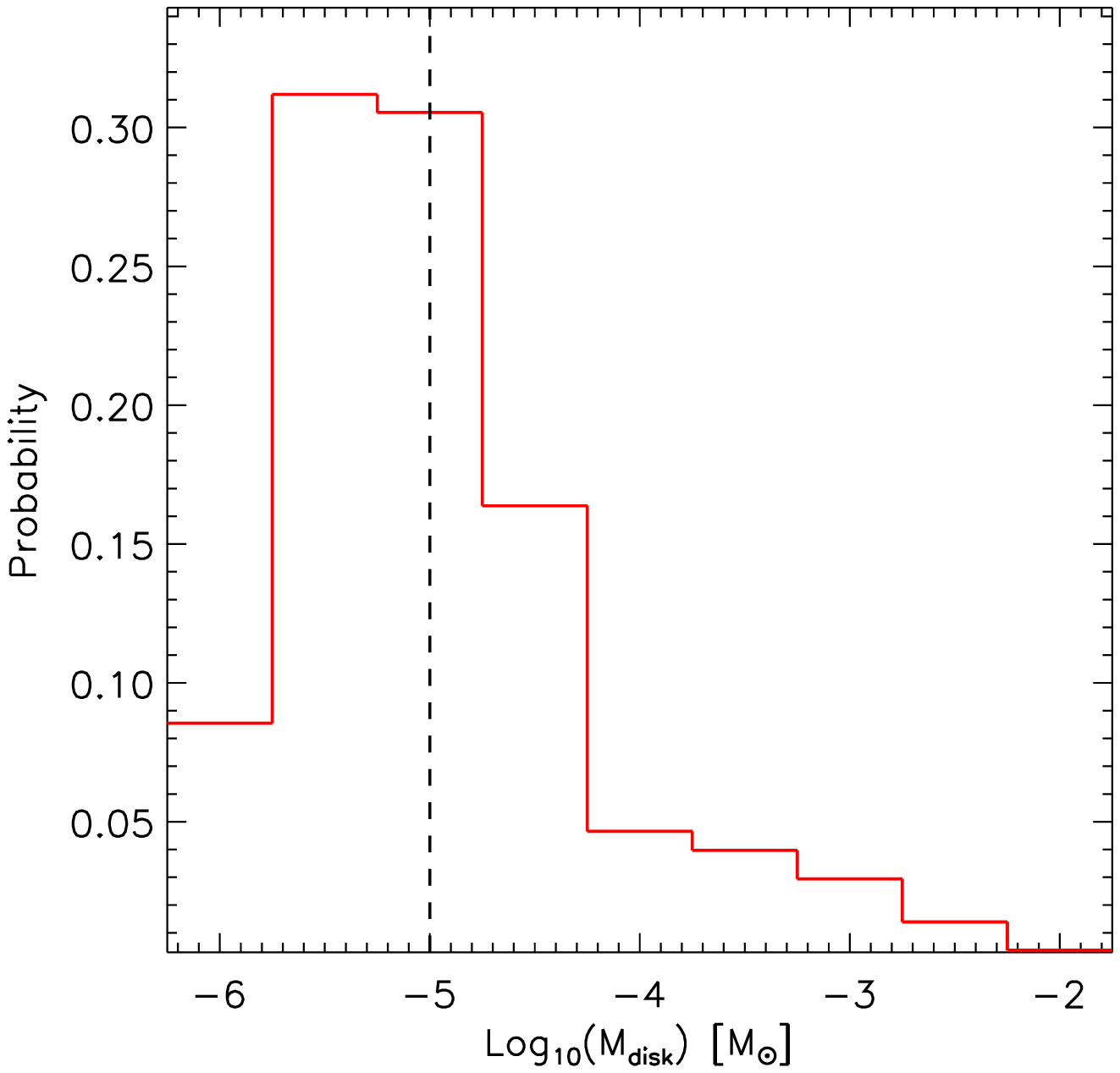}
\end{minipage}
\begin{minipage}[c]{0.4\textwidth}
\centering
\includegraphics[width=2.1in]{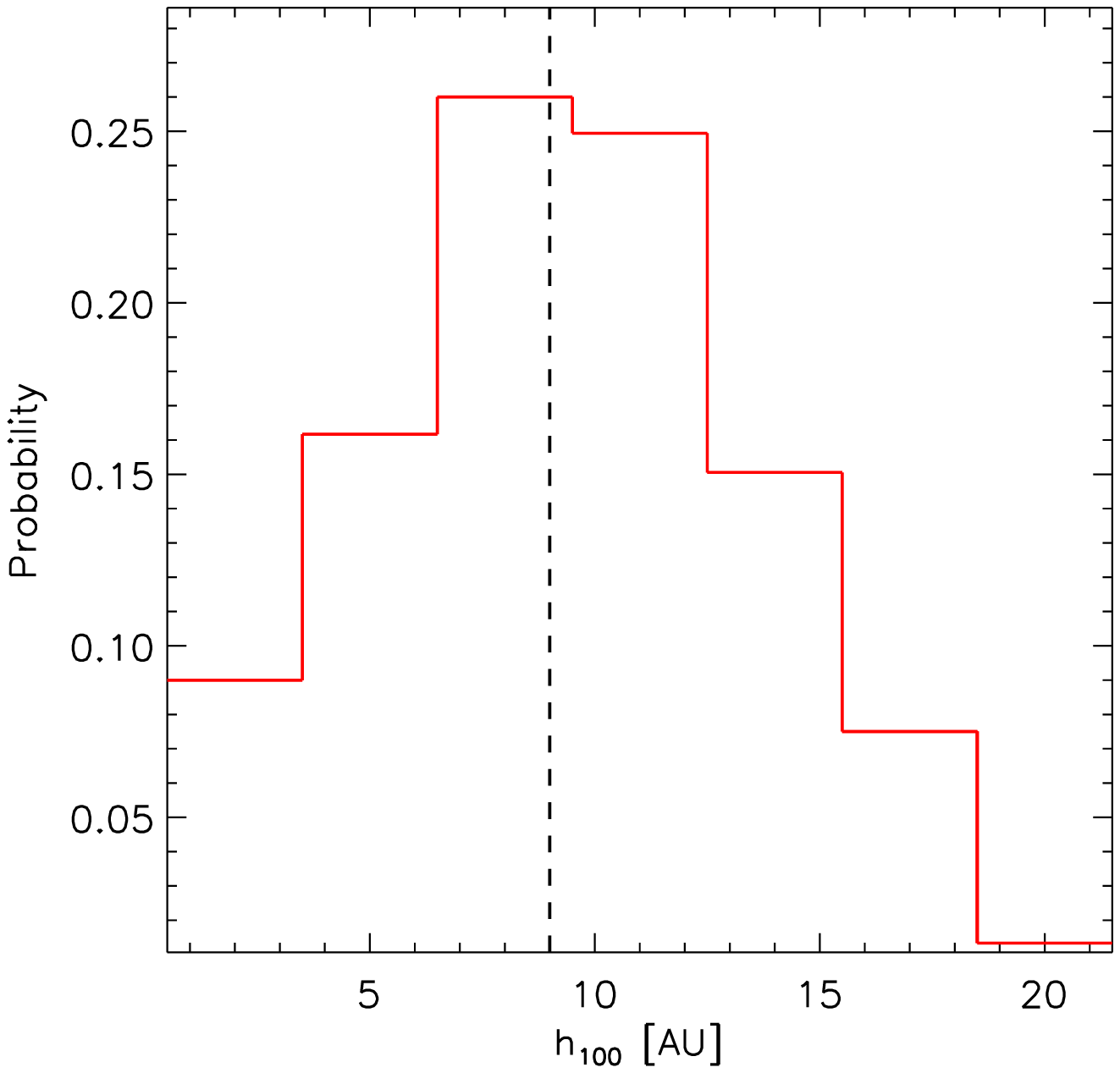}
\includegraphics[width=2.1in]{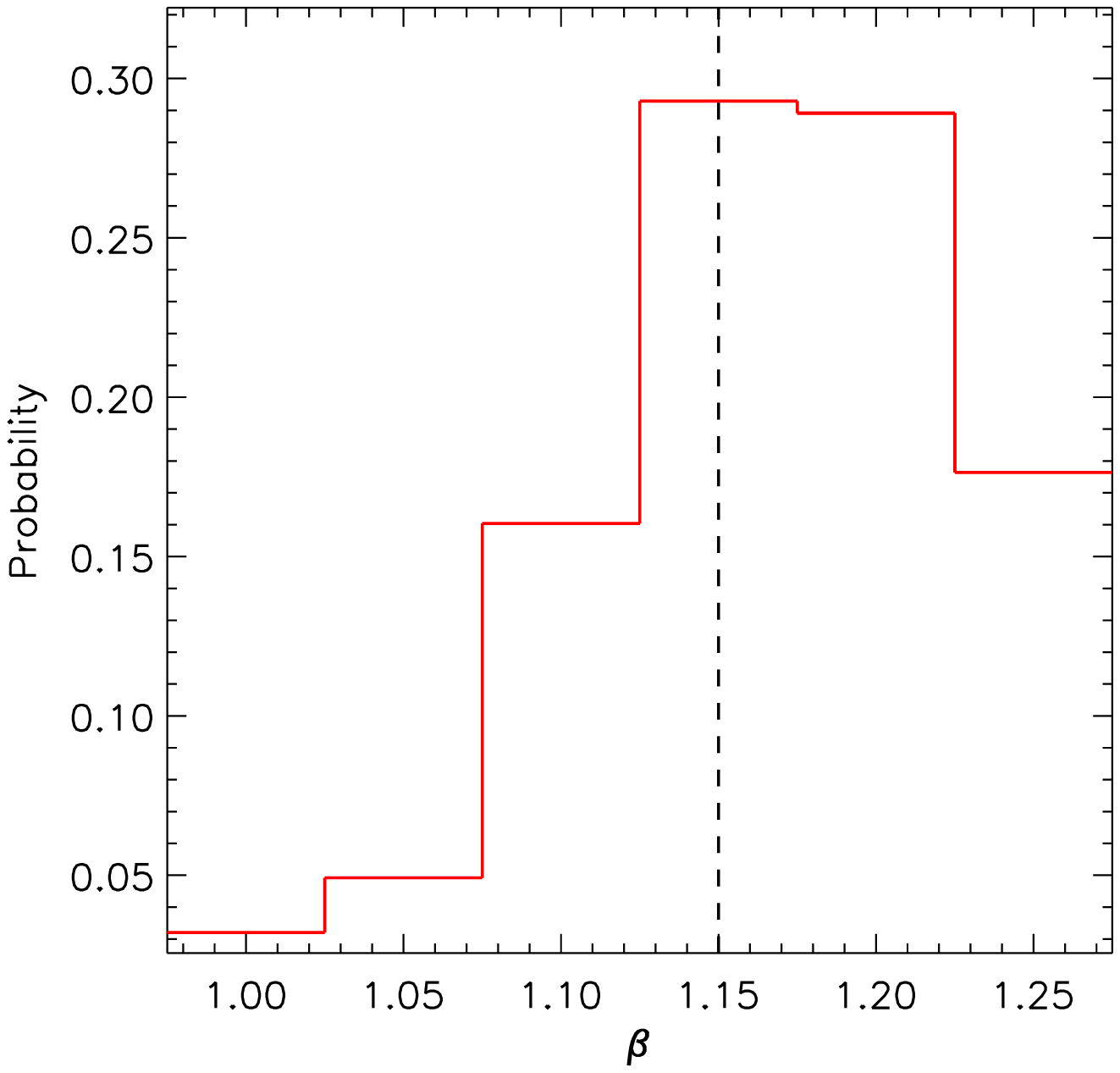}
\end{minipage}
\begin{minipage}[c]{0.4\textwidth}
\centering
\includegraphics[width=2.1in]{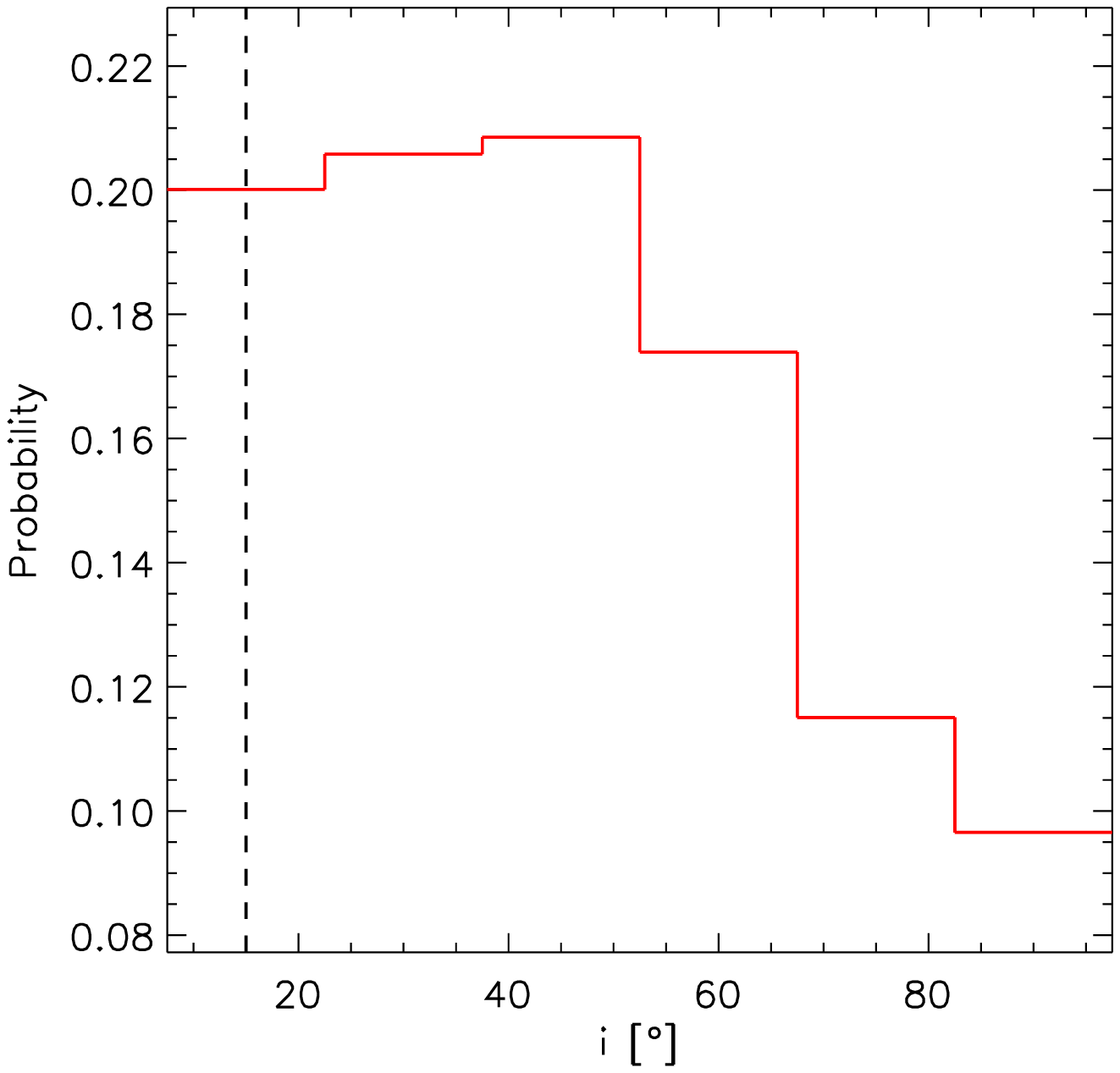}
\end{minipage}
\caption{Probability distributions of model parameters for TWA 32. The vertical dashed lines denote the best-fit values.}
\end{figure*}

\newpage
\begin{figure*}[t]
\centering
\begin{minipage}[c]{0.4\textwidth}
\centering
\includegraphics[width=2.1in]{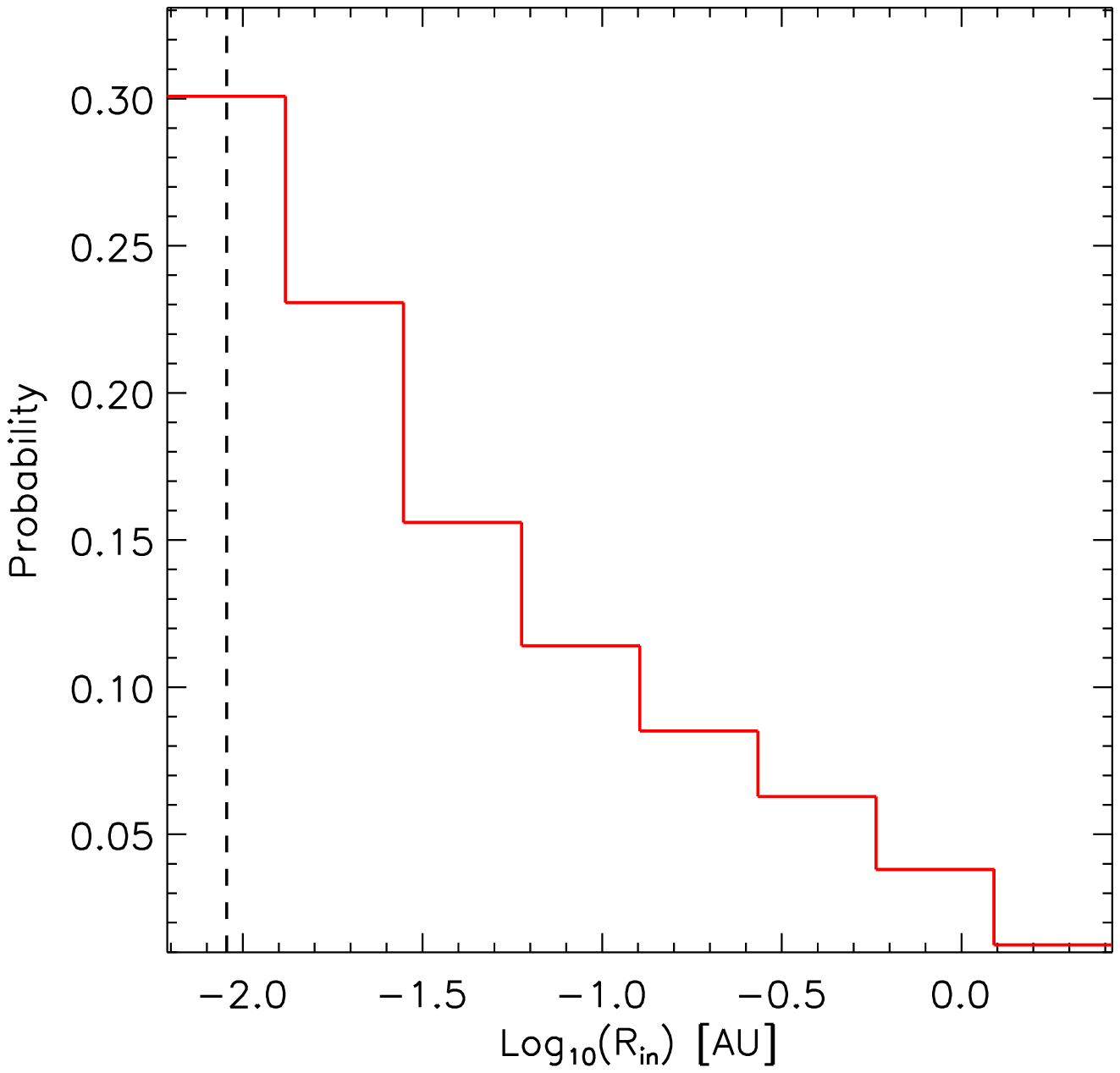}
\includegraphics[width=2.1in]{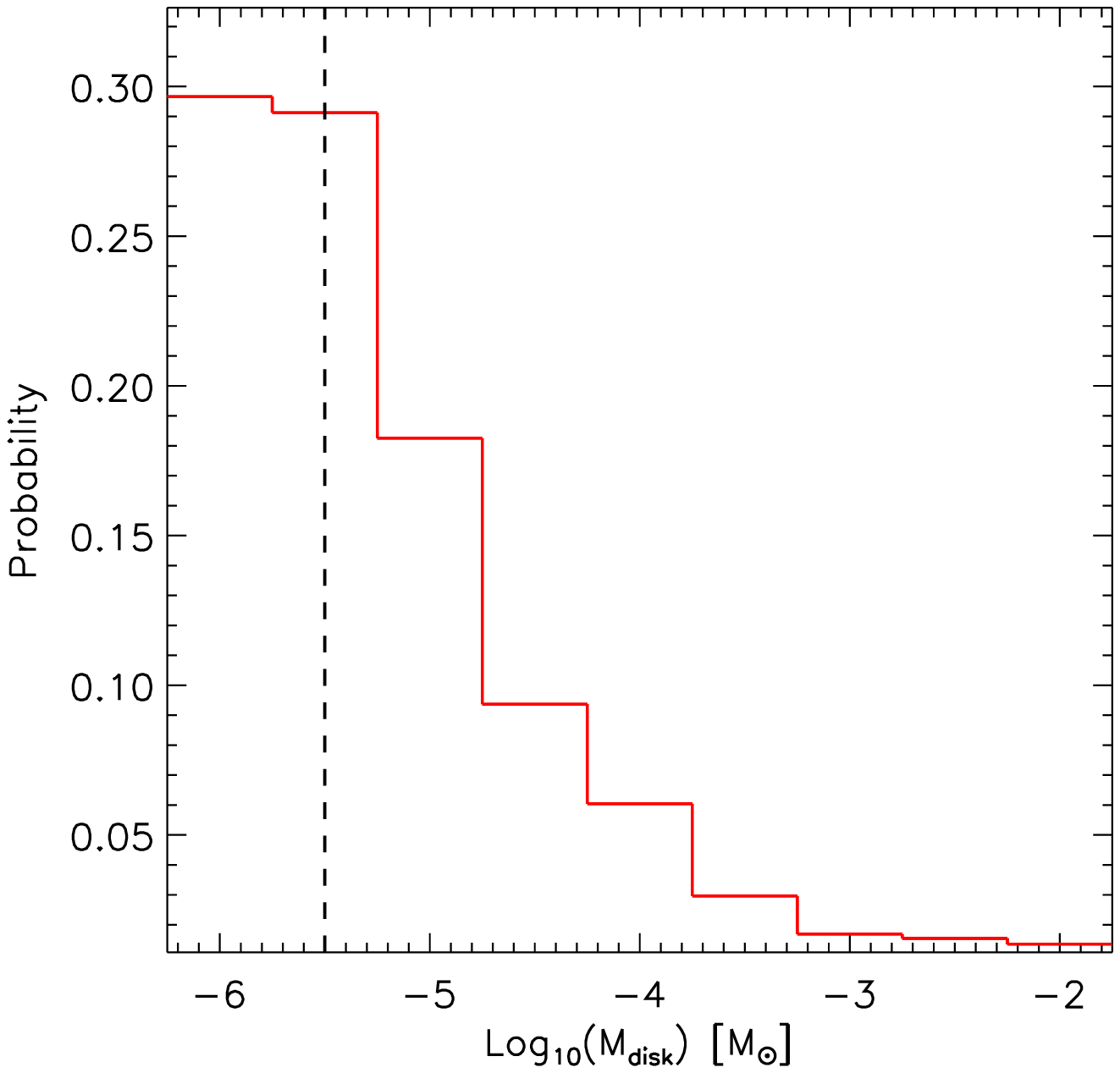}
\end{minipage}
\begin{minipage}[c]{0.4\textwidth}
\centering
\includegraphics[width=2.1in]{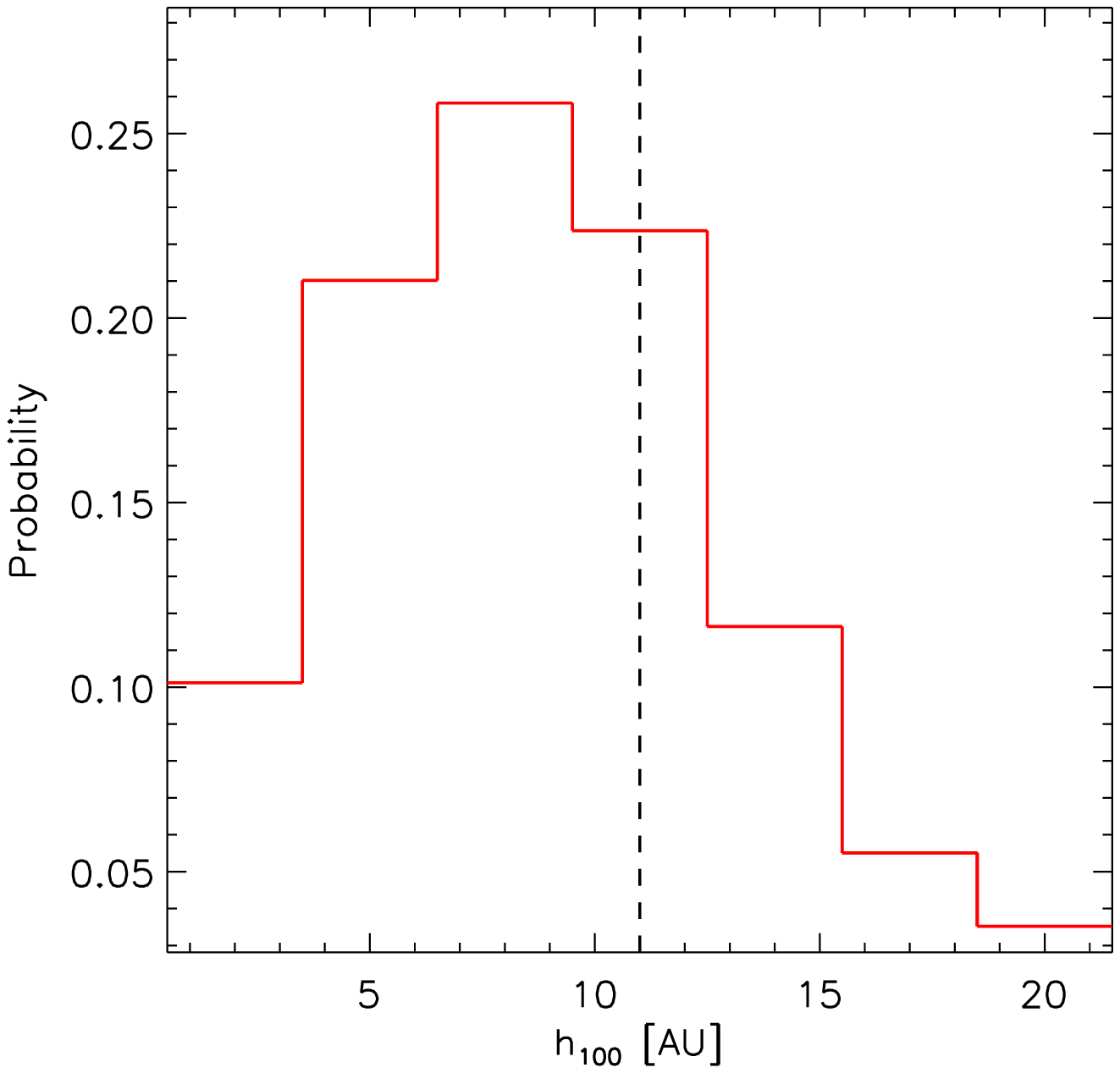}
\includegraphics[width=2.1in]{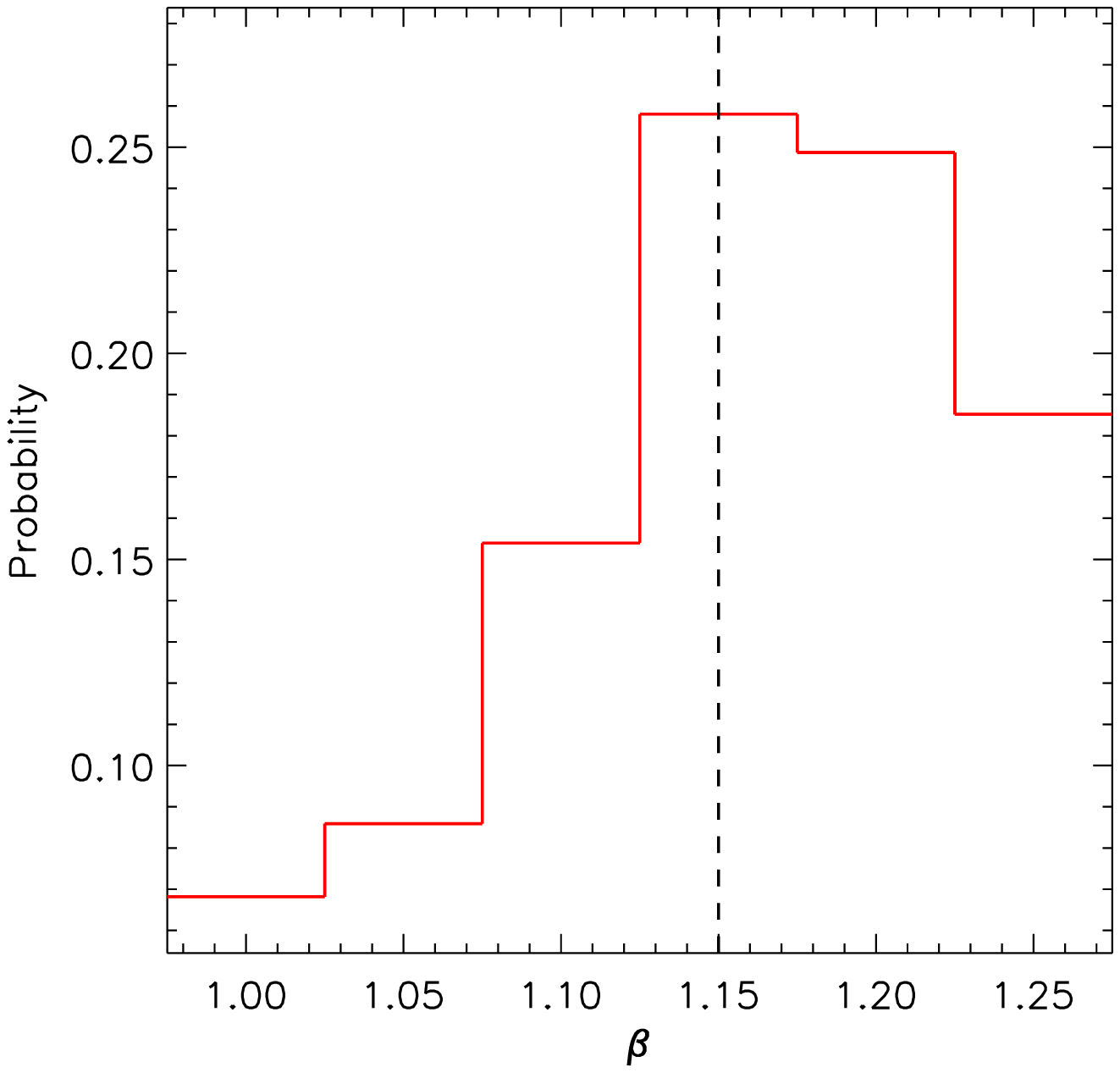}
\end{minipage}
\begin{minipage}[c]{0.4\textwidth}
\centering
\includegraphics[width=2.1in]{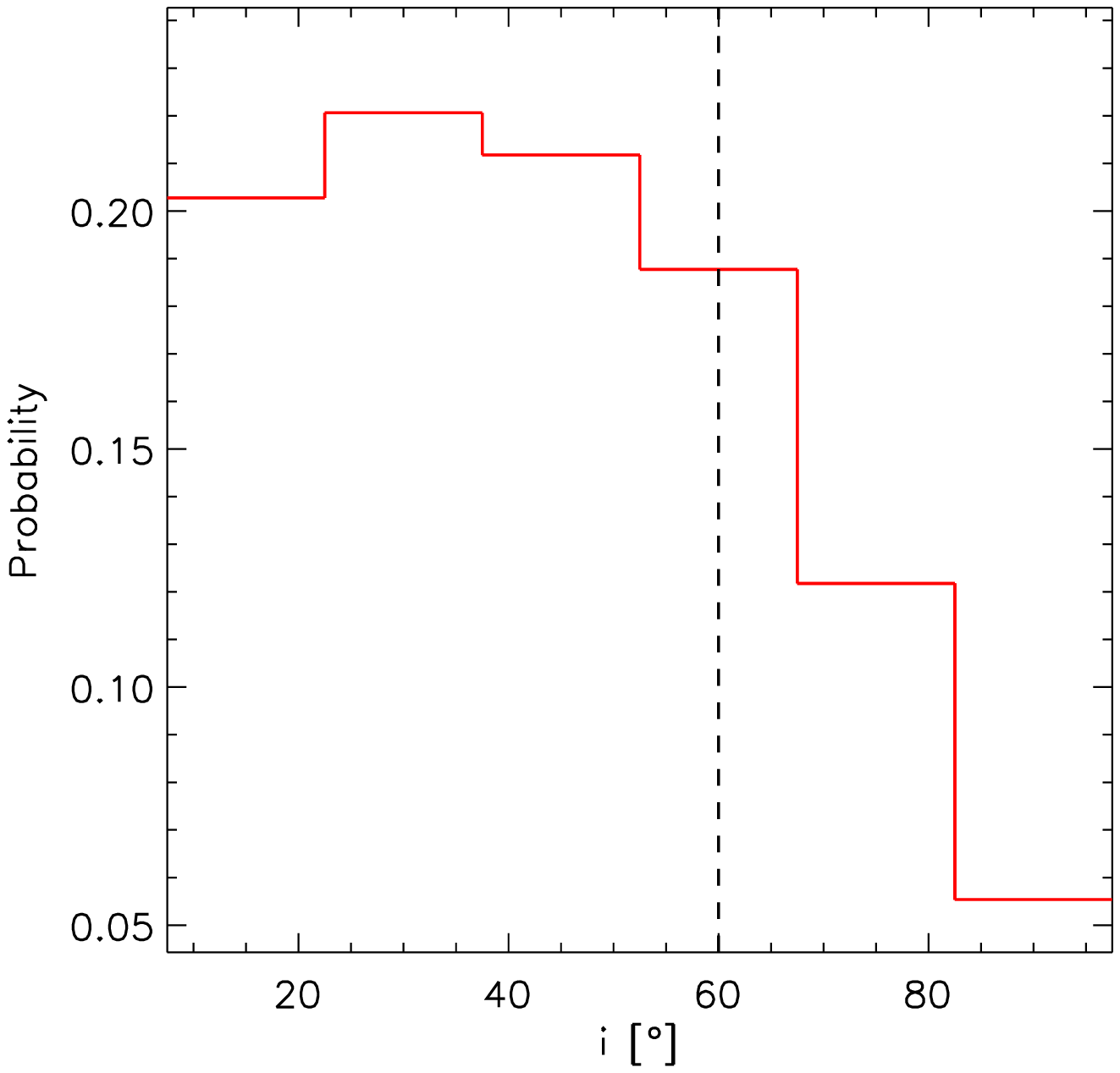}
\end{minipage}
\caption{Probability distributions of model parameters for TWA 34. The vertical dashed lines denote the best-fit values.}
\end{figure*}
\end{appendix}

\end{document}